\documentclass[preprint,amssymb,nobibnotes,nofootinbib,aps,prc]{revtex4}
\usepackage{psfig}
\usepackage{graphicx}
\usepackage{bm}
\newcommand{\bea}{\begin{eqnarray}}
\newcommand{\eea}{\end{eqnarray}}
\newcommand{\be}{\begin{equation}}
\newcommand{\ee}{\end{equation}}
\newcommand{\bt}{\begin{tabular}}
\newcommand{\et}{\end{tabular}}
\newcommand{\Tr}{{\rm Tr}}
\newcommand{\no}{\nonumber}
\newcommand{\ovl}{\overline}

\newcommand{\Si}{ \mbox{\boldmath $\Sigma$}  }

\newcommand{\pa}{\partial}
\newcommand{\beas}{\begin{eqnarray*}}
\newcommand{\eeas}{\end{eqnarray*}}

\newcommand{\fr}{\frac}

\newcommand{\dg}{\dagger}
\newcommand{\La}{\Lambda}
\newcommand{\pam}{\partial_\mu}
\begin{document}

\title{Kaons and antikaons in hot and dense hadronic matter}

\author{A.Mishra}
\email{mishra@th.physik.uni-frankfurt.de}
\affiliation{Institut f\"ur Theoretische Physik,
     J.W. Goether Universit\"at,
        Robert Mayer Str. 8-10, D-60054 Frankfurt am Main, Germany}

\author{E.L. Bratkovskaya}
\email{Elena.Bratkovskaya@th.physik.uni-frankfurt.de}
\affiliation{Institut f\"ur Theoretische Physik,
     J.W. Goether Universit\"at,
        Robert Mayer Str. 8-10, D-60054 Frankfurt am Main, Germany}

\author{J. Schaffner-Bielich}
\affiliation{Institut f\"ur Theoretische Physik,
     J.W. Goether Universit\"at,
        Robert Mayer Str. 8-10, D-60054 Frankfurt am Main, Germany}

\author{S. Schramm}
\affiliation{Institut f\"ur Theoretische Physik,
     J.W. Goether Universit\"at,
        Robert Mayer Str. 8-10, D-60054 Frankfurt am Main, Germany}

\author{H.~St\"ocker}
\affiliation{Institut f\"ur Theoretische Physik,
     J.W. Goether Universit\"at,
        Robert Mayer Str. 8-10, D-60054 Frankfurt am Main, Germany}

\begin{abstract}
The medium modification of kaon and antikaon masses, compatible with
low energy KN scattering data, are studied in a chiral SU(3) model.
The mutual interactions with baryons in hot hadronic matter and the
effects from the baryonic Dirac sea on the K(${\rm {\bar K}}$) masses
are examined. The in-medium masses from the chiral SU(3) effective
model are compared to those from chiral perturbation theory.
Furthermore, the influence of these in-medium effects on kaon rapidity
distributions and transverse energy spectra as well as the $K, \bar{K}$
flow pattern in heavy-ion collision experiments at 1.5 to 2 A$\cdot$GeV
are investigated within the HSD transport approach. Detailed
predictions on the transverse momentum and rapidity dependence of
directed flow $v_1$ and the elliptic flow $v_2$ are provided for Ni+Ni
at 1.93 A$\cdot$GeV within the various models, that can be used to
determine the in-medium $K^\pm$ properties from the experimental side
in the near future.
\end{abstract}


\maketitle

\section{Introduction}

The property of hadrons under extreme conditions of temperature and
density  \cite{QM02} is an important topic in present strong
interaction physics.  This subject has direct implications in heavy-ion
collision experiments, in the study of astrophysical compact objects
(like neutron stars) as well as in the early universe.  The in-medium
properties of kaons have been primarily investigated due to their
relevance in neutron star phenomenology as well as relativistic
heavy-ion collisions. For example, in the interior of the neutron star
the attractive kaon nucleon interaction might lead to kaon condensation
as suggested early by Kaplan and Nelson \cite{kaplan}.  The in-medium
modification of kaon/antikaon properties can be observed experimentally
primarily in relativistic nuclear collisions.  Indeed, the experimental
\cite{FOPI,Laue99,kaosnew,Sturm01,Forster02} and theoretical studies
\cite{lix,cmko,Li2001,Cass97,brat97,CB99,laura03,Effenber00,Aichelin,Fuchs}
on $K^\pm$ production from A+A collisions at SIS energies of 1-2
A$\cdot$GeV have shown that in-medium properties of kaons have been
seen in the collective flow pattern of $K^+$ mesons as well as in the
abundancy and spectra of antikaons.

The theoretical research work on the topic of in-medium properties of
hadrons was triggered in part by the early suggestion of Brown and Rho
\cite{brown}, that the modifications of hadron masses should scale with
the scalar quark condensate $\langle q\bar{q}\rangle$ at finite baryon
density.  The first attempts to extract the antikaon-nucleus potential
from the analysis of kaonic-atom data were in favor of very strong
attractive potentials of the order of -150 to -200 MeV at normal
nuclear matter density $\rho_0$ \cite{FGB94,Gal}.  However, more recent
self-consistent calculations based on a chiral Lagrangian
\cite{Lutz98,Lutz021,Lutz02,Oset00} or coupled-channel G-matrix theory
(within meson-exchange potentials) \cite{lauran} only predicted
moderate attractive depths of -50 to -80 MeV at density $\rho_0$.

The problem with the antikaon potential at finite baryon density is
that the antikaon-nucleon amplitude in the isospin channel $I=0$ is
dominated by the $\Lambda(1405)$ resonant structure, which in free
space is only 27 MeV below the ${\bar K}N$ threshold. It is presently
not clear if this physical resonance is a real excited state of a
'strange' baryon or it is some short lived intermediate state which
can be generated dynamically in a coupled channel
$T$-matrix scattering equation using a suitable meson-baryon potential.
Additionally, the coupling between the ${\bar K}N$ and $\pi Y$
($Y=\Lambda,\Sigma$) channels is essential to get the proper dynamical
behavior in free space. Correspondingly, the in-medium properties of
the $\Lambda(1405)$, such as its pole position and its width, which in
turn influence strongly the antikaon-nucleus optical potential, are
very sensitive to the many-body treatment of the medium effects.
Previous works have shown that a self-consistent treatment of the
$\bar{K}$ self energy has a strong impact on the scattering amplitudes
\cite{Lutz98,Oset00,Laura,Effenber00,Lutz02,lauran} and thus on the
in-medium properties of the antikaon.  Due to the complexity of this
many-body problem the actual kaon and antikaon self energies (or
potentials) are still a matter of debate.

In the present investigation, we will use a chiral SU(3) model for the
description of hadrons in the medium \cite{paper3}.  The nucleons -- as modified in
the hot hyperonic matter -- have been studied within this model
\cite{kristof1} previously. Furthermore, the properties of vector mesons
\cite{hartree,kristof1} -- due to their interactions  with nucleons in
the medium -- have been also examined and have been found to have
appreciable modifications due to Dirac sea polarization effects. The
chiral SU(3)$_{flavor}$ model was also been generalized to
SU(4)$_{flavor}$ to study the mass modification of D-mesons due to
their interactions with the light hadrons in hot hadronic matter
in \cite{dmeson}. In the present work, the masses of kaons (antikaons),
as modified in the medium due to their interaction with nucleons, are
studied within the chiral SU(3) framework, which is consistent with the
low energy KN scattering data \cite{juergen}. In this approach, however,
only the real parts of the kaon/antikaon self energies can be
addressed.

Another problem related to the complexity of kaon/antikaon physics in
relativistic heavy-ion reactions is that not only the mean-field
properties of the $K, \bar{K}$ mesons enter, but also their in-medium
scattering amplitudes sometimes far from the mass shell
\cite{lauran,laura03}, because the
antikaon couples strongly to the baryons and achieves a nontrivial
spectral width in the medium. Whereas in the early calculations in
Refs.  \cite{lix,cmko,Li2001,Cass97,brat97,CB99} the off-shell
transition amplitudes have been simply extrapolated from on-shell cross
sections in vacuum, more recent studies in Ref. \cite{laura03} have
incorporated the full off-shell dynamics in transport. Thus, when
confronting our model predictions with experimental data, we will use a
parametrization of the off-shell amplitudes from Ref.
\cite{lauran,laura03} in order to reduce the systematic uncertainty in
the transition amplitudes or scattering rates of antikaons with
baryons. For the real part of the self energies, however, we will
implement the results from the different models to be discussed in this
work.

The outline of the paper is as follows: In section II we shall briefly
review the chiral SU(3) model used in the present investigation.
Section III describes the medium modification of the K($\bar K$) mesons
in this effective model. In section IV, we investigate the kaon masses
using chiral perturbation theory, while in section V we discuss
and compare the results from the chiral SU(3) model to those from
the chiral perturbation theory.
Section VI explores the effects of in-medium modifications of kaons
(antikaons) on their production and flow pattern in relativistic
heavy-ion collisions in comparison to experimental data.  Section VII
summarizes the findings of the present investigation and discusses
future extensions.

\section{ The hadronic chiral $SU(3) \times SU(3)$ model }
In this section the various terms of the effective hadronic Lagrangian
used
\be
{\cal L} = {\cal L}_{kin} + \sum_{ W =X,Y,V,{\cal A},u }{\cal L}_{BW}
          + {\cal L}_{VP} + {\cal L}_{vec} + {\cal L}_0 + {\cal L}_{SB}
\label{genlag}
\ee
are discussed. Eq. (\ref{genlag}) corresponds to a relativistic quantum
field theoretical model of baryons and mesons built on
a nonlinear realization of chiral symmetry
and broken scale invariance (for details see \cite{paper3,hartree,kristof1})
to describe strongly interacting nuclear matter.
The model was used successfully to describe nuclear matter, finite nuclei,
hypernuclei and neutron stars.
The Lagrangian contains the baryon octet, the spin-0 and spin-1 meson
multiplets as the elementary degrees of freedom. In Eq. (\ref{genlag}),
$ {\cal L}_{kin} $ is the kinetic energy term, $  {\cal L}_{BW}  $
contains the baryon-meson interactions in which the baryon-spin-0 meson
interaction terms generate the baryon masses. $ {\cal L}_{VP} $
describes the interactions of vector mesons with the pseudoscalar
mesons (and with photons).  $ {\cal L}_{vec} $ describes the dynamical
mass generation of the vector mesons via couplings to the scalar
mesons and contains additionally quartic self-interactions of the
vector fields.  ${\cal L}_0 $ contains the meson-meson interaction terms
inducing the spontaneous breaking of chiral symmetry as well as
a scale invariance breaking logarithmic potential. $ {\cal L}_{SB} $
describes the explicit chiral symmetry breaking.

The kinetic energy terms are given as
\bea
\label{kinetic}
{\cal L}_{kin} &=& i\Tr \overline{B} \gamma_{\mu} D^{\mu}B
                + \frac{1}{2} \Tr D_{\mu} X D^{\mu} X
+  \Tr (u_{\mu} X u^{\mu}X +X u_{\mu} u^{\mu} X)
                + \frac{1}{2}\Tr D_{\mu} Y D^{\mu} Y \nonumber \\
               &+&\frac {1}{2} D_{\mu} \chi D^{\mu} \chi
                - \frac{ 1 }{ 4 } \Tr
\left(\tilde V_{ \mu \nu } \tilde V^{\mu \nu }  \right)
- \frac{ 1 }{ 4 } \Tr \left(F_{ \mu \nu } F^{\mu \nu }  \right)
- \frac{ 1 }{ 4 } \Tr \left( {\cal A}_{ \mu \nu } {\cal A}^{\mu \nu }
 \right)\, .
\eea
In (\ref{kinetic}) $B$ is the baryon octet, $X$ the scalar meson
multiplet, $Y$ the pseudoscalar chiral singlet, $\tilde{V}^\mu$ (${\cal
A}^\mu$) the renormalised vector (axial vector) meson multiplet with
the field strength tensor
$\tilde{V}_{\mu\nu}=\pa_\mu\tilde{V}_\nu-\pa_\nu\tilde{V}_\mu$ $({\cal
A}_{\mu\nu}= \pa_\mu{\cal A}_\nu-\pa_\nu{\cal A}_\mu $), $F_{\mu\nu}$
is the field strength tensor of the photon and $\chi$
is the scalar, iso-scalar dilaton (glueball) -field.
In the above, $u_\mu= -\fr{i}{2}[u^\dg\pam u - u\pam u^\dg]$,
where $u=\exp\Bigg[\fr{i}{\sigma_0}\pi^a\lambda^a\gamma_5\Bigg]$
is the unitary transformation operator, and the covariant derivative
reads $ D_\mu = \pam\, + [\Gamma_\mu,\,\,]$, with
$\Gamma_\mu=-\fr{i}{2}[u^\dg\pam u + u\pam u^\dg]$.

The baryon -meson interaction for a general meson field $W$
has the form
\be
{\cal L}_{BW} =
-\sqrt{2}g_8^W \left(\alpha_W[\ovl{B}{\cal O}BW]_F+ (1-\alpha_W)
[\ovl{B} {\cal O}B W]_D \right)
- g_1^W \frac{1}{\sqrt{3}} \Tr(\ovl{B}{\cal O} B)\Tr W  \, ,
\ee
with $[\ovl{B}{\cal O}BW]_F:=\Tr(\ovl{B}{\cal O}WB-\ovl{B}{\cal O}BW)$ and
$[\ovl{B}{\cal O}BW]_D:= \Tr(\ovl{B}{\cal O}WB+\ovl{B}{\cal O}BW)
- \frac{2}{3}\Tr (\ovl{B}{\cal O} B) \Tr W$.
The different terms -- to be considered -- are those for the interaction
of baryons  with
scalar mesons ($W=X, {\cal O}=1$), with
vector mesons  ($W=\tilde V_{\mu}, {\cal O}=\gamma_{\mu}$ for the vector and
$W=\tilde V_{\mu \nu}, {\cal O}=\sigma^{\mu \nu}$ for the tensor
interaction),
with axial vector mesons ($W={\cal A}_\mu, {\cal O}=\gamma_\mu \gamma_5$)
and with
pseudoscalar mesons ($W=u_{\mu},{\cal O}=\gamma_{\mu}\gamma_5$), respectively.
For the current investigation the following interactions are relevant:
Baryon-scalar meson interactions generate the baryon masses through
coupling of the baryons to the non-strange $ \sigma (\sim
\langle\bar{u}u + \bar{d}d\rangle) $ and the strange $
\zeta(\sim\langle\bar{s}s\rangle) $ scalar quark condensate.
The parameters $ g_1^S$, $g_8^S $ and $\alpha_S$ are adjusted
to fix the baryon masses to their experimentally measured
vacuum values. It should be emphasised that the nucleon mass
also depends on the {\em strange condensate} $ \zeta $.
For the special case of ideal mixing ($\alpha_S=1$ and $g_1^S=\sqrt 6 g_8^S$)
the nucleon mass depends only on the non--strange quark condensate.
In the present investigation, the general case will be used
to study hot and strange hadronic matter \cite{kristof1}, which
takes into account the baryon coupling terms to both scalar fields
($\sigma$ and $\zeta$) while summing over the baryonic tadpole diagrams to
investigate the effect from the baryonic Dirac sea in the relativistic
Hartree approximation \cite{kristof1}.

In analogy to the baryon-scalar meson coupling there exist two
independent baryon-vector meson interaction terms corresponding
to the F-type (antisymmetric) and $D$-type (symmetric) couplings.
Here we will use the antisymmetric coupling because -- from
the universality principle  \cite{saku69} and
the vector meson dominance model -- one can conclude
that the symmetric coupling should be small.
We realize it by setting $\alpha_V=1$
for all fits. Additionally we decouple the strange vector field
$ \phi_\mu\sim\bar{s}
\gamma_\mu s $ from the nucleon by setting $ g_1^V=\sqrt{6}g_8^V $.
The remaining baryon-vector meson interaction reads
\be
{\cal L}_{BV}=-\sqrt{2}g_8^V\Big\{[\bar{B}\gamma_\mu BV^\mu]_F+\Tr\big(\bar{B}\gamma_\mu B\big)
\Tr V^\mu\Big\}\, .
\ee

The Lagrangian describing the interaction for the scalar mesons, $X$,
and pseudoscalar singlet, $Y$, is given as \cite{paper3}
\bea
\label{cpot}
{\cal L}_0 &= &  -\frac{ 1 }{ 2 } k_0 \chi^2 I_2
     + k_1 (I_2)^2 + k_2 I_4 +2 k_3 \chi I_3,
\eea
with $I_2= \Tr (X+iY)^2$, $I_3=\det (X+iY)$ and $I_4 = \Tr (X+iY)^4$.
In the above, $\chi$ is the scalar color singlet gluon field. It is
introduced in order to 'mimic' the QCD trace anomaly, i.e. the nonvanishing
energy-momentum tensor $\theta_\mu^\mu = (\beta_{QCD}/2g)\langle
G^a_{\mu\nu}G^{a,\mu\nu}\rangle$, where $G^a_{\mu\nu}$ is the gluon
field tensor.
A scale breaking potential is introduced:
\be
\label{lscale}
  {\cal L}_{\mathrm{scalebreak}}=- \frac{1}{4}\chi^4 \ln
   \frac{ \chi^4 }{ \chi_0^4}
 +\frac{\delta}{3}\chi^4 \ln \frac{I_3}{\det \langle X \rangle_0}
\ee
which allows for the identification of the $\chi$ field width the gluon
condensate $\theta_\mu^\mu=(1-\delta)\chi^4$.
Finally the term
${\cal L}_{\chi} = - k_4 \chi^4 $
generates a phenomenologically consistent finite vacuum expectation
value. The variation of
$\chi$ in the medium is rather small \cite{paper3}.
Hence we shall use the frozen glueball approximation i.e. set
$\chi$ to its vacuum value, $\chi_0$.

The Lagrangian for the vector meson interaction is written as
\bea
{\cal L}_{vec} &=&
    \fr{m_V^2}{2}\fr{\chi^2}{\chi_0^2}\Tr\big(\tilde{V}_\mu\tilde{V}^\mu\big)
+   \fr{\mu}{4}\Tr\big(\tilde{V}_{\mu\nu}\tilde{V}^{\mu\nu}X^2\big) 
+ \fr{\lambda_V}{12}\Big(\Tr\big(\tilde{V}^{\mu\nu}\big)\Big)^2 +
    2(\tilde{g}_4)^4\Tr\big(\tilde{V}_\mu\tilde{V}^\mu\big)^2  \, .
\eea
The vector meson fields, $\tilde{V}_\mu$ are related to the
renormalized fields by
$V_\mu = Z_V^{1/2}\tilde{V}_\mu$, with $V = \omega, \rho, \phi \, $.
The masses of $\omega,\rho$ and $\phi$ are fitted from $m_V, \mu$ and
$\lambda_V$.

The explicit symmetry breaking term is given as \cite{paper3}
\be
 {\cal L}_{SB}=\Tr A_p\left(u(X+iY)u+u^\dagger(X-iY)u^\dagger\right)
\label{esb-gl}
\ee
with $A_p=1/\sqrt{2}{\mathrm{diag}}(m_{\pi}^2 f_{\pi},m_\pi^2 f_\pi, 2 m_K^2 f_K
-m_{\pi}^2 f_\pi)$ and $m_{\pi}=139$ MeV, $m_K=498$ MeV. This
choice for $A_p$, together with the constraints
$\sigma_0=-f_\pi$, $\zeta_0=-\frac {1}{\sqrt 2} (2 f_K -f_\pi)$
on the VEV on the scalar condensates assure that
the PCAC-relations of the pion and kaon are fulfilled.
With $f_{\pi} = 93.3$~MeV and $f_K = 122$~MeV we obtain $|\sigma_0| =
93.3$~MeV and $|\zeta_0 |= 106.56$~MeV.

\subsection{Mean field approximation}

We proceed to study the hadronic properties in the chiral SU(3) model.
The Lagrangian density in the mean field approximation is given as
\begin{eqnarray}
{\cal L}_{BX}+{\cal L}_{BV} &=& -\sum_i\overline{\psi_{i}}\, [g_{i
\omega}\gamma_0 \omega + g_{i\phi}\gamma_0 \phi
+m_i^{\ast} ]\,\psi_{i} \\
{\cal L}_{vec} &=& \frac{1}{2}m_{\omega}^{2}\frac{\chi^2}{\chi_0^2}\omega^
2+g_4^4 \omega^4 +
\frac{1}{2}m_{\phi}^{2}\frac{\chi^2}{\chi_0^2}\phi^2+g_4^4
\left(\fr{Z_\phi}{Z_\omega}\right)^2\phi^4\\
{\cal V}_0 &=& \frac{ 1 }{ 2 } k_0 \chi^2 (\sigma^2+\zeta^2)
- k_1 (\sigma^2+\zeta^2)^2
     - k_2 ( \frac{ \sigma^4}{ 2 } + \zeta^4)
     - k_3 \chi \sigma^2 \zeta \nonumber \\
&+& k_4 \chi^4 + \frac{1}{4}\chi^4 \ln \frac{ \chi^4 }{ \chi_0^4}
 -\frac{\delta}{3} \chi^4 \ln \frac{\sigma^2\zeta}{\sigma_0^2 \zeta_0} \\
{\cal V}_{SB} &=& \left(\frac{\chi}{\chi_0}\right)^{2}\left[m_{\pi}^2 f_{\pi}
\sigma
+ (\sqrt{2}m_K^2 f_K - \frac{ 1 }{ \sqrt{2} } m_{\pi}^2 f_{\pi})\zeta
\right],
\end{eqnarray}
where $m_i^* = -g_{\sigma i}{\sigma}-g_{\zeta i}{\zeta} $ is the
effective mass of the baryon of type i ($i = N, \Si, \La, \Xi $).
In the above, $g_4=\sqrt {Z_\omega} \tilde g_4$ is the renormalised
coupling for $\omega$-field.
The thermodynamical potential of the grand
canonical ensemble $\Omega$ per unit volume $V$ at given chemical
potential $\mu$ and temperature $T$ can be written as
\bea
\frac{\Omega}{V} &=& -{\cal L}_{vec} - {\cal L}_0 - {\cal L}_{SB}
- {\cal V}_{vac} + \sum_i\frac{\gamma_i }{(2 \pi)^3}
\int d^3k\,
E^{\ast}_i(k)\Big(f_i(k)+\bar{f}_i(k)
\Big)\nonumber \\
&&- \sum_i\frac{\gamma_i }{(2 \pi)^3}\,\mu^{\ast}_i
\int d^3k\,\Big(f_i(k)-\bar{f}_i(k)\Big).
\label{OmegaV}
\eea
Here the vacuum energy (the potential at $\rho=0$) has been subtracted
in order to get a vanishing vacuum energy. In (\ref{OmegaV}) $\gamma_i$
are the spin-isospin degeneracy factors.  The $f_i$ and $\bar{f}_i$ are
thermal distribution functions for the baryon of species $i$, given in
terms of the effective single particle energy, $E^\ast_i$, and chemical
potential, $\mu^\ast_i$, as
\bea
f_i(k) &=& \fr{1}{{\rm e}^{\beta (E^{\ast}_i(k)-\mu^{\ast}_i)}+1}\quad ,\quad
\bar{f}_i(k)=\fr{1}{{\rm e}^{\beta (E^{\ast}_i(k)+\mu^{\ast}_i)}+1}, \no
\eea
with $E^{\ast}_i(k) = \sqrt{k_i^2+{m^\ast_i}^2}$ and $ \mu^{\ast}_i
                = \mu_i-g_{i\omega}\omega$.
The mesonic field equations are determined by minimizing the
thermodynamical potential \cite{hartree,kristof1}.
These are expressed in terms of
the scalar and vector densities
for the baryons at finite temperature
\bea
\rho^s_i = \gamma_i
\int \frac{d^3 k}{(2 \pi)^3} \,\frac{m_i^{\ast}}{E^{\ast}_i}\,
\left(f_i(k) + \bar{f}_i(k)\right) \, ; \;\;
\rho_i = \gamma_i \int \frac{d^3 k}{(2 \pi)^3}\,\left(f_i(k) -
\bar{f}_i(k)\right) \,.
\label{dens}
\eea
The energy density and the pressure are given as,
$\epsilon = \Omega/V+\mu_i\rho_i $+TS and $ p = -\Omega/V $.

\subsection{Relativistic Hartree approximation}
%
%
The relativistic Hartree approximation takes into account the effects from
the Dirac sea by summing over the baryonic tadpole diagrams.
The dressed propagator for a baryon of type $i$ has the form
\cite{vacpol}
\begin{eqnarray}
 && G_i^H(p) = \left(\gamma^\mu\bar{p}_\mu+m_i^\ast\right)
\Bigg[\frac{1}{\bar{p}^2 -{m_i^\ast}^2+i\epsilon}\nonumber\\
&+&\frac{\pi i}{E_i^\ast(p)}\left\{\frac{\delta(\bar{p}^0
-E_i^\ast(p))}{{\rm e}^{\beta(E_i^\ast(p)-\mu_i^\ast)}+1}
+ \frac{\delta(\bar{p}^0+E_i^\ast(p))}{
{\rm e}^{\beta(E_i^\ast(p)+\mu_i^\ast)}+1} \right\}\Bigg]\nonumber \\
&\equiv & G_i^F(p) + G_i^D(p),
\end{eqnarray}
where $E_i^\ast(p)=\sqrt{{\bf p}^2+{m_i^\ast}^2}$, $\bar{p}=p+\Sigma_i^V$
and $m_i^\ast=m_i+\Sigma_i^S$. $\Sigma_i^V$ and $\Sigma_i^S$
are the vector and scalar self energies of baryon, $i$ respectively.
In the present investigation (for the study of hot baryonic matter)
the baryons couple to both the non-strange ($\sigma$) and strange
($\zeta$) scalar fields, so that we have
\be
\Sigma^S_i=-(g_{\sigma i}\tilde{\sigma}+g_{\zeta
i}\tilde{\zeta})\, ,
\ee
where $\tilde{\sigma}=\sigma-\sigma_0$, $\tilde{\zeta}=\zeta-\zeta_0$.
The scalar self-energy $\Sigma^S_i$ can be written
\bea
\Sigma^S_i= i\left(\fr{g_{\sigma i}^2}{m_\sigma^2}
 + \fr{g_{\zeta i}^2}{m_\zeta^2}\right)\int\fr{\rm{d}^4p}{(2\pi)^4}
 \Tr\big[G_i^F(p)+G_i^D(p)\big]e^{ip^0\eta}
\equiv \big(\Sigma^S_i\big)^F + \big(\Sigma^S_i\big)^D \, .
\eea
$ (\Sigma^S_i)^D $ is the density dependent part and is identical
to the mean field contribution
\be
\big(\Sigma^S_i\big)^D = -\left(\fr{g_{\sigma i}^2}{m_\sigma^2}
+ \fr{g_{\zeta i}^2}{m_\zeta^2}\right)\rho_i^s,
\label{denpart}
\ee
with $\rho_i^s$ as defined in (\ref{dens}).
The Feynman part $ (\Sigma^S_i)^F $ of the scalar part of the self-energy
is divergent. We carry out a dimensional regularization to extract
the convergent part.
Adding the counter terms \cite{kristof1}
\be
\label{ctc}
\left(\Sigma^S_i\right)_{CTC} = - \left(\fr{g_{\sigma i}^2}{m_\sigma^2}
+\fr{g_{\zeta i}^2}{m_\zeta^2}\right)\sum_{n=0}^3\fr{1}{n!}
(g_{\sigma i}\tilde{\sigma}+g_{\zeta i}\tilde{\zeta})^n\beta_{n+1}^{i}\, ,
\ee
yields the additional contribution from the Dirac sea to
the baryon self energy \cite{kristof1}.
The field equations for the scalar meson fields are then modified to
\be
\frac{\partial(\Omega/V)}{\partial\Phi}\Bigg|_{RHA} =
\frac{\partial(\Omega/V)}{\partial\Phi}\Bigg|_{MFT}
+\sum_i\frac{\pa m_i^\ast}{\pa\Phi}\Delta \rho^s_i = 0
\quad\mbox{with}\quad \Phi = \sigma, \zeta\, ,
\ee
where the additional contribution to the nucleon scalar density is
given as \cite{kristof1}
\be
\Delta\rho^s_i = -\fr{\gamma_i}{4\pi^2} \left[ {m_i^\ast}^3\ln\left(
                  \fr{m_i^\ast}{m_i}\right)
+ m_i^2(m_i-m_i^\ast) - \fr{5}{2}m_i(m_i-m_i^\ast)^2 + \fr{11}{6}
  (m_i-m_i^\ast)^3\right].
\ee

\section{Kaon interactions in the effective chiral model}
\label{kmeson}

We now examine the medium modification for the $K$-meson mass
in hot and dense hadronic matter. In the last section,
the SU(3) chiral model was used to study the hadronic properties
in the medium within the relativistic Hartree approximation.
In this section, we investigate the medium modification
of the $K$-meson mass due to the interactions
of the $K$-mesons in the hadronic medium.

In the chiral effective model as used here,
the interactions to the scalar fields (nonstrange, $\sigma$ and strange,
$\zeta$) as well as a vectorial interaction and a $\omega$- exchange
term modify the masses for K$^\pm$ mesons in the medium.
These interactions were considered within the SU(3) chiral model
to investigate the modifications of K-mesons in thermal
medium \cite{kmeson} in the mean field approximation.
The scalar meson exchange gives an attractive
interaction leading to a drop of the K -meson masses similar
to a scalar sigma term in the chiral perturbation theory \cite{kaplan}.
In fact, the KN \cite{kmeson} as well as the $\pi N$ sigma term
are predicted in our approach automatically by using SU(3)
symmetry. The pion-nucleon and kaon-nucleon sigma terms
as calculated  from the scalar meson exchange interaction of
our Lagrangian are 28 MeV and 463 MeV, respectively. The value
for KN sigma term calculated in our model is close to the value
of $\Sigma_{KN}$=450 MeV found by lattice gauge calculations
\cite{ksgn}. In addition to the terms considered in \cite{kmeson},
we also account the effect of repulsive scalar contributions
($\sim (\partial _\mu K^+)(\partial ^\mu K^-)$) which contribute
in the same order as the attractive sigma term in chiral perturbation
theory. These terms will ensure that KN scattering lengths can be
described and, hence, the low density theorem for kaons is fulfilled.

The scalar meson multiplet has the expectation value
$\langle X \rangle
= \rm {diag}( \sigma/\sqrt 2, \sigma/\sqrt 2 , \zeta ) $,
with $\sigma$ ans $\zeta$ corresponding to the non-strange and strange
scalar condensates.
The pseudoscalar meson field P can be written as,
\begin{equation}
P = \left(
\begin{array}{ccc} \pi^0/\sqrt 2 & \pi^+ & \frac{2 K^+}{1+w} \\
\pi^- & -\pi^0/\sqrt 2 & 0 \\
\frac {2 K^-}{1+w} & 0 & 0
\\ \end{array}\right),
\end{equation}
where $w=\sqrt 2 \zeta/\sigma$ and we have written down
the entries which are relevant for the present investigation.
From PCAC, one gets the decay constants for the pseudoscalar mesons
as $f_\pi=-\sigma$ and $f_K=-(\sigma +\sqrt 2 \zeta )/2$.
The vector meson interaction with
the pseudoscalar mesons, which modifies the masses of the K mesons,
is given as \cite {kmeson}
\begin{equation}
{\cal L} _ {VP}= -\frac{m_V^2}{2g_V} {\rm {Tr}} (\Gamma_\mu V^\mu) +
{\rm  h.c.}
\label{lvp}
\end{equation}
The vector meson multiplet is given as
$V = {\rm  diag}\big ((\omega +\rho_0)/\sqrt 2,\;
(\omega -\rho_0)/\sqrt 2,\; \phi\big )$. The non-diagonal
components in the multiplet, which are not relevant in the present
investigation, are not written down. With the interaction (\ref{lvp}),
the coupling of the $K$-meson to the $\omega$-meson is related to  the
pion-rho coupling as
$g_{\omega K}/g_{\rho \pi \pi}=f_\pi^2 /(2f_K^2)$.

The scalar meson exchange interaction term, which is attractive
for the $K$-mesons, is given from the explicit symmetry breaking term
by equation (\ref {esb-gl}), where $A_p =1/\sqrt 2$
diag ($m_\pi^2 f_\pi$, $m_\pi^2 f_\pi$, 2 $m_K^2 f_K -m_\pi^2 f_\pi$).

The interaction Lagrangian modifying the $K$-meson mass can be written
as \cite{kmeson}
\begin{eqnarray}
\cal L _{K} & = & -\frac {3i}{8 f_K^2} \bar N \gamma^\mu N
( K^- \partial_\mu K ^+ - \partial_\mu K ^- K^+)\nonumber \\
 &+ & \frac{m_K^2}{2f_K} (\sigma +\sqrt 2 \zeta) K^- K^+
-i g_{\omega K} ( K^- \partial_\mu K ^+ - \partial_\mu K ^- K^+)
\omega ^ \mu \nonumber \\
& - & \frac {1}{f_K} (\sigma +\sqrt 2 \zeta)
(\partial _\mu K^-)(\partial ^\mu K^+)
+\frac {d_1}{2 f_K^2}(\bar N N)
(\partial _\mu K^-)(\partial ^\mu K^+).
\label{lagd}
\end{eqnarray}
In (\ref{lagd}) the first term is the vectorial interaction term
obtained from the first term in (\ref{kinetic}) (Weinberg-Tomozawa
term).  The second term, which gives an attractive interaction for the
$K$-mesons, is obtained from the explicit symmetry breaking term
(\ref{esb-gl}). The third term, referring to the interaction in terms
of $\omega$-meson exchange, is attractive for the $K^-$ and repulsive
for $K^+$.  The fourth term arises within the present chiral model from
the kinetic term of the pseudoscalar mesons given by the third term in
equation (\ref{kinetic}), when the scalar fields in one of the meson
multiplets, $X$ are replaced by their vacuum expectation values.  The
fifth term in (\ref{lagd}) for the KN interactions arises from the term
\begin{equation}
{\cal L }^{BM} =d_1 Tr (u_\mu u ^\mu \bar B B),
\label{dtld}
\end{equation}
in the SU(3) chiral model. The last two terms in (\ref{lagd})
represent the range term in the chiral model.
From the Fourier transformation of the equation-of-motion for kaons
$$-\omega^2 + m_K^2 +\Sigma_K(\omega,\rho)=0$$ one derives the effective
energy of the $K^+$ and $K^-$ which are the poles of the kaon propagator in
the medium (assuming zero momentum for S-wave Bose condensation) where
$\Sigma_K$ denotes the kaon selfenergy in the medium.

\subsection{Fitting to KN scattering data}

For the $KN$ interactions, the term (\ref{dtld}) reduces to
\begin{equation}
{\cal L }_{\tilde D }^{KN} =d_1 \frac{1}{2 f_K^2}
(\bar N N) (\partial _\mu K^-) (\partial ^\mu K^+).
\end{equation}
The coefficient $d_1$ in the above is determined by fitting to the
$KN$ scattering length \cite{thorsson,barnes,juergen,dmeson}.  The
isospin averaged $KN$ scattering length
\be
\bar a _{KN}= \frac {1}{4} (3 a_{KN}^{I=1}+ a_{KN}^{I=0})
\ee
can be calculated to be
\bea
\bar a _{KN} &=&\frac {m_K}{4\pi (1+m_K/m_N)} \Big [
-\big(\frac {m_K}{2 f_K} \big )\cdot \frac {g_{\sigma N}}{m_\sigma^2}
-\big(\frac {\sqrt 2 m_K}{2 f_K} \big )\cdot \frac {g_{\zeta N}}{m_\zeta^2}
\nonumber \\
 & - & \frac {2 g_{\omega K} g_{\omega N}}{m_\omega ^2}
-\frac {3}{4 f_K^2} + \frac {d_1 m_K}{2 f_K^2} \Big ].
\eea
The empirical value of the isospin averaged scattering length
\cite{thorsson,juergen,barnes} is taken to be
\be
\bar a _{KN} \approx -0.255 ~ \rm {fm}
\label{aknemp}
\ee
which determines the value for the coefficient $d_1$.
The present calculations use the values,
$g_{\sigma N}=10.618,\;\; {\rm and}\;\; g_{\zeta N}=-0.4836$ as
fixed by the vacuum baryon masses,
and the other parameters are fitted to the nuclear matter
saturation properties as listed in Ref. \cite{kristof1}.
We consider the case when a quartic vector interaction is present.
The coefficient $d_1$ is evaluated in the mean field and RHA cases
as ${5.63}/{m_K}$ and ${4.33}/{m_K}$ respectively \cite{dmeson}.
The contribution from this term is thus seen to be attractive,
contrary to the other term proportional to
$(\partial_\mu K^-)(\partial ^\mu K^+)$ in (\ref{lagd})
which is repulsive.

\section{Chiral perturbation theory}

The effective Lagrangian obtained from chiral perturbation theory
\cite{kaplan} has been used extensively in the literature for the study
of kaons in dense matter. This approach has a vector interaction
(called the Tomozawa-Weinberg term) as the leading term.
At sub-leading order there are the attractive scalar nucleon
interaction term (the sigma term) \cite{kaplan} as well as
the repulsive scalar contribution (proportional to the kinetic
term of the pesudoscalar meson).
The $KN$ interaction is given as
\begin {eqnarray}
{\cal L}_ {KN} & = &
-\frac {3i}{8 f_K^2} \bar N \gamma^\mu N
( K^- \partial_\mu K ^+ - \partial_\mu K ^- K^+)
+ \frac{\Sigma_{KN}}{f_K^2} (\bar N N ) K^- K^+\nonumber \\
&+ & \frac {\tilde D}{f_K^2}
(\bar N N) (\partial _\mu K^-) (\partial ^\mu K^+).
\label{ldcpt}
\end{eqnarray}
where $\Sigma _{KN} =\frac{\bar m +m_s}{2}
\langle N | (\bar u u + \bar s s) | N \rangle$ \cite{ksgn}.
In our calculations, we take $m_s=150$~MeV and $\bar m=(m_u+m_d)/2=7$~MeV.

The last term of the Lagrangian (\ref{ldcpt}) is repulsive and is
of the same order as the attractive sigma term. This, to a large
extent, compensates the scalar attraction due to the scalar
$\Sigma$- term. The coefficient $\tilde D$ is fixed by
the $KN$ scattering lengths (see ref.\cite{thorsson}) by
choosing a value for $\Sigma_{KN}$,
which depends on the strangeness content of the nucleon.
Its value has, however, a large uncertainty.
We consider the two extreme choices:
$\Sigma _{KN}= 2 m_\pi$ and $\Sigma_{K N}= 450$~MeV.
The coefficient, $\tilde {D}$ as fitted to the empirical
value of the KN scattering length (\ref{aknemp}) is
in general given by \cite{thorsson}
\begin{equation}
\tilde D \approx 0.33/m_K - \Sigma_{KN} /m_K^2.
\end{equation}

In the next section, we shall discuss the results for
the $K$-meson mass modification obtained in the effective
chiral model as compared to chiral perturbation theory.

\section{Medium modification of K-meson masses}
\label{kmass}

We now investigate the $K$-meson masses in hot and dense hadronic
medium within a chiral SU(3) model.  The contributions from the various
terms of the interaction Lagrangian (\ref{lagd}) are shown in Fig. 1 in
the mean field approximation.  The vector interaction as well as
the $\omega$ exchange terms (given by the first and the third terms
of equation (\ref{lagd}), respectively) lead to a drop for the $K^-$
mass, whereas they are repulsive for the $K^+$. The scalar meson
exchange term is attractive for both $K^+$ and $K^-$. The first
term of the range term of eq. (\ref{lagd}) is
repulsive whereas the second term has an attractive contribution.  This
results in a turn over of the $K$-mass at around 0.8 $\rho_0$ above
which the attractive range term (the last term in (\ref{lagd}))
 dominates.  The dominant contributions arise from the scalar exchange
and the range term (dominated by $d_1$ term at higher densities),
which lead to a substantial drop of K meson mass in the medium. The
vector terms lead to a further drop of $K^-$ mass, whereas
for $K^+$ they compete with the contributions from the other two
contributions.  The effect from the nucleon Dirac sea on the mass
modification of the $K$-mesons is shown in Fig. 2.  This gives rise to
smaller modifications as compared to the mean field calculations though
qualitative features remain the same.

\begin{figure}
\begin{center}
\centerline{\parbox[b]{8cm}{
\includegraphics[width=9.2cm,height=9cm]{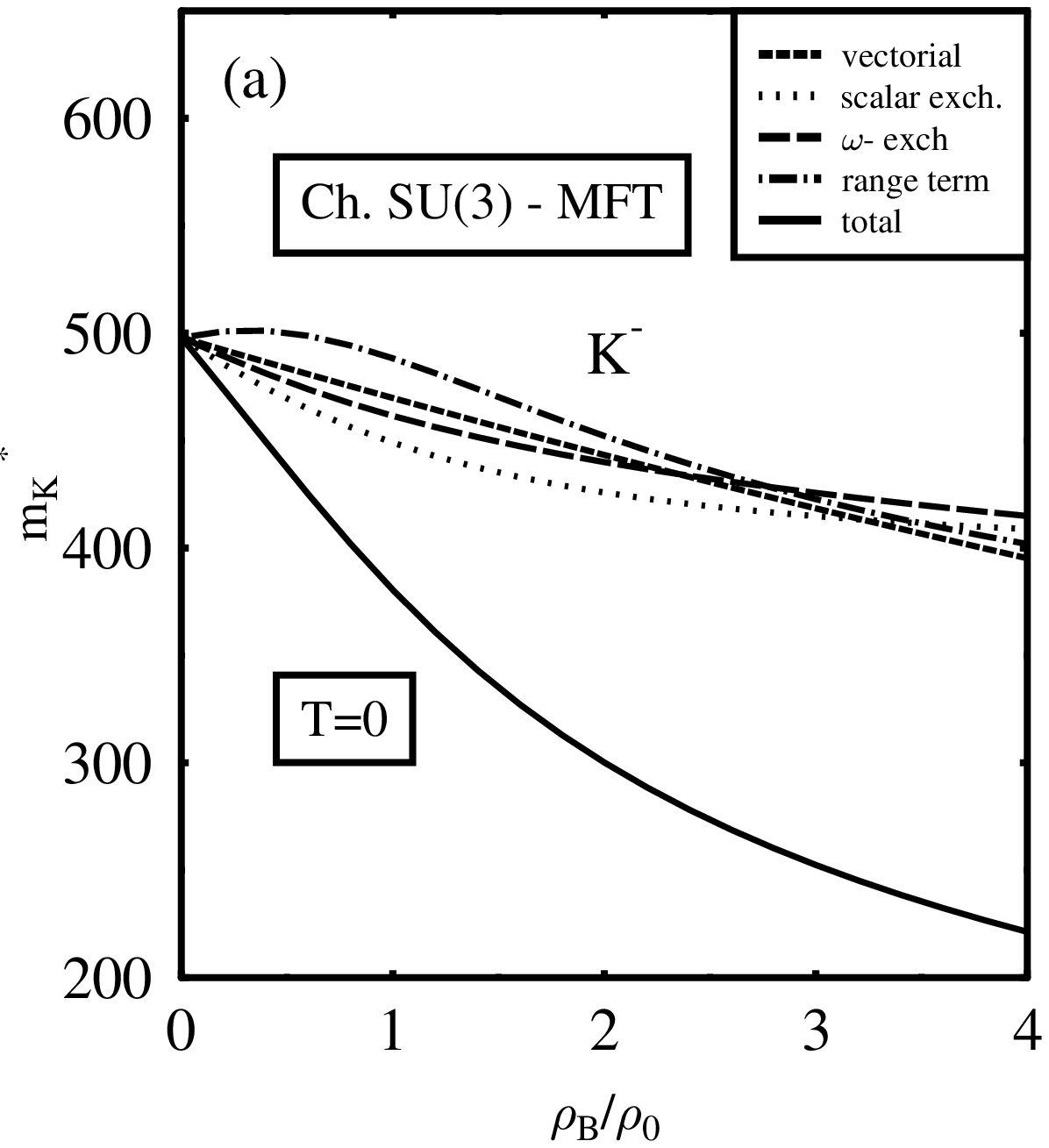}}
\parbox[b]{8cm}{
\includegraphics[width=9.2cm,height=9cm]{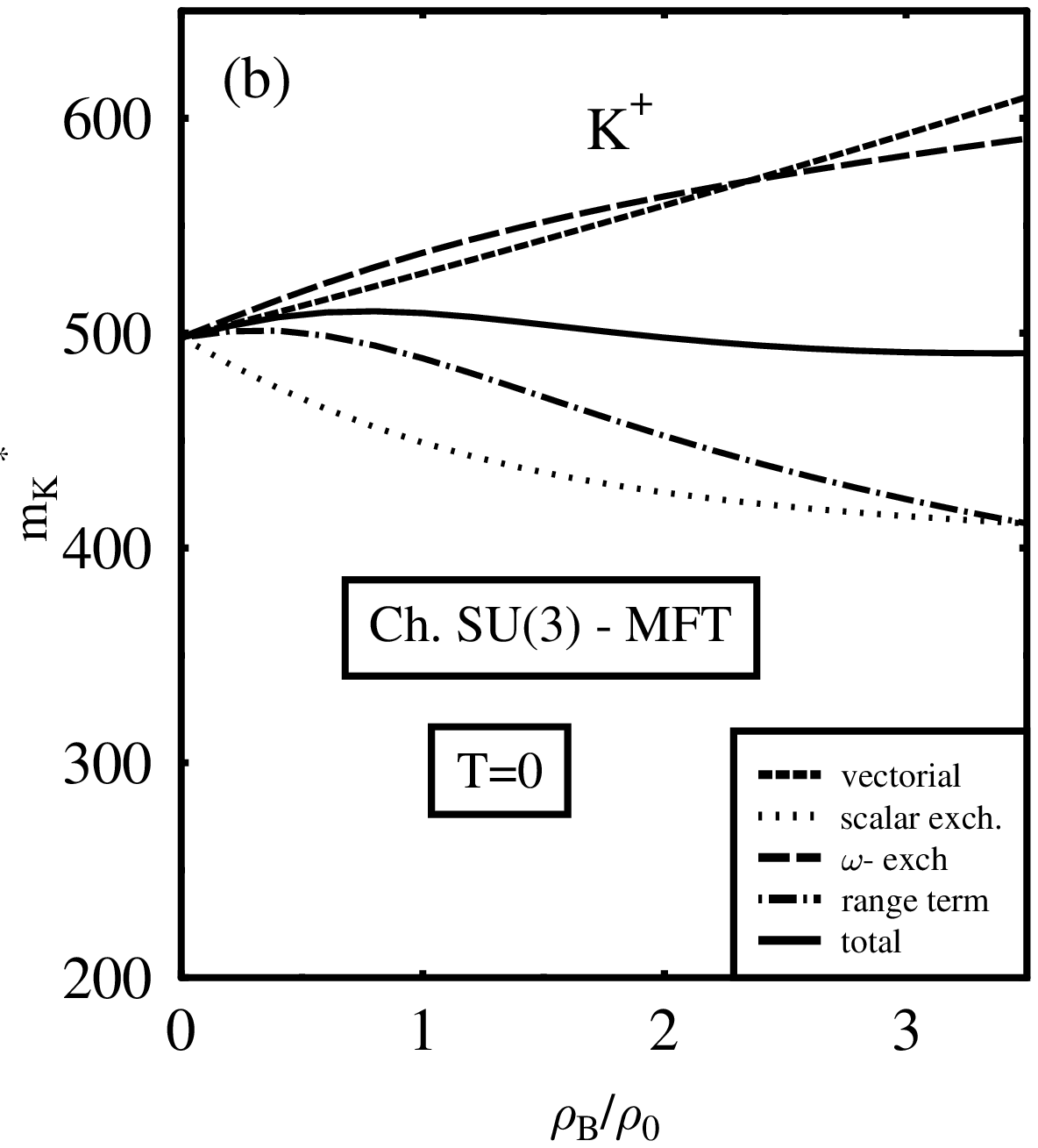}}}
\caption{
\label{mkt0imft}
Contributions to the masses of K$^\pm$ mesons due to the various
interactions in the effective chiral model in the mean field
approximation. The curves refer to individual contributions from
the vectorial interaction, scalar exchange, $\omega$ exchange,
the range term.
The solid line shows to the total contribution.}
\end{center}
\end{figure}

\begin{figure}
\begin{center}
\centerline{\parbox[b]{8cm}{
\includegraphics[width=9.2cm,height=9cm]{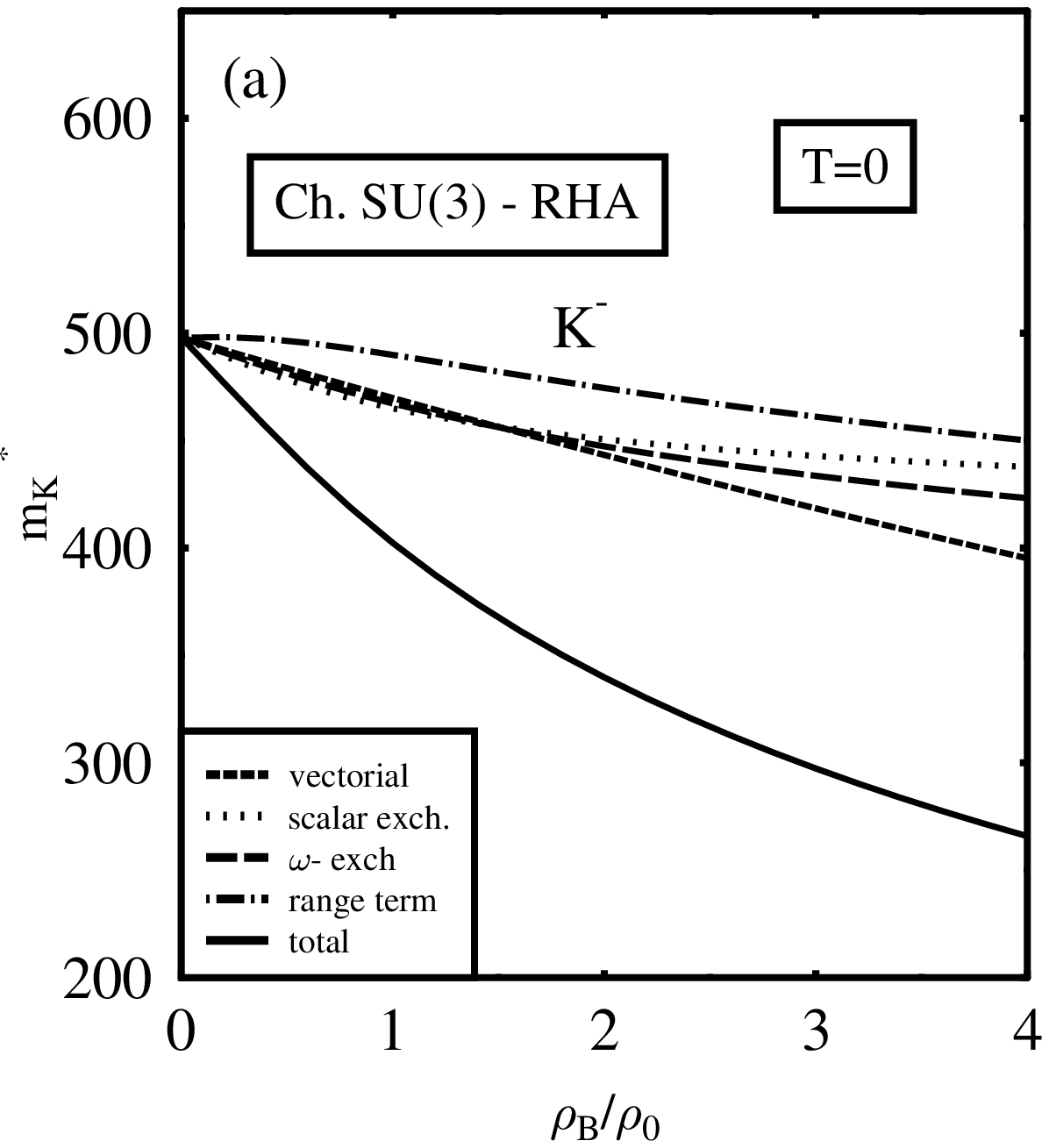}}
\parbox[b]{8cm}{
\includegraphics[width=9.2cm,height=9cm]{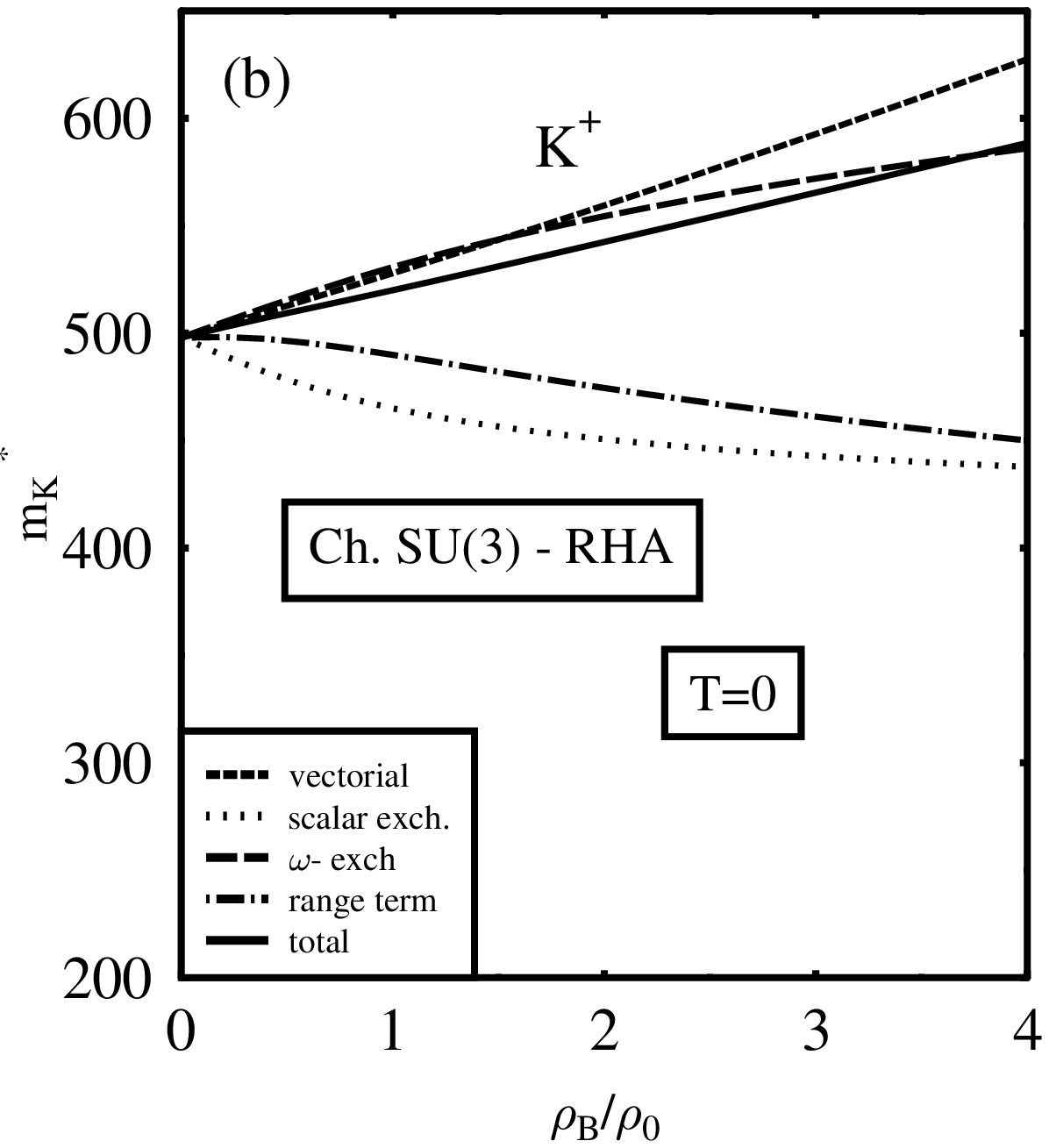}}}
\caption{
\label{mkt0irha}
Same as in figure (\protect\ref{mkt0imft}), but in the relativistic
Hartree approximation. The contributions to the masses of $K^\pm$ mesons
due to the various interactions are seen to be smaller with the
Dirac sea effects.}
\end{center}
\end{figure}

\begin{figure}
\begin{center}
\centerline{\parbox[b]{8cm}{
\includegraphics[width=9.2cm,height=9cm]{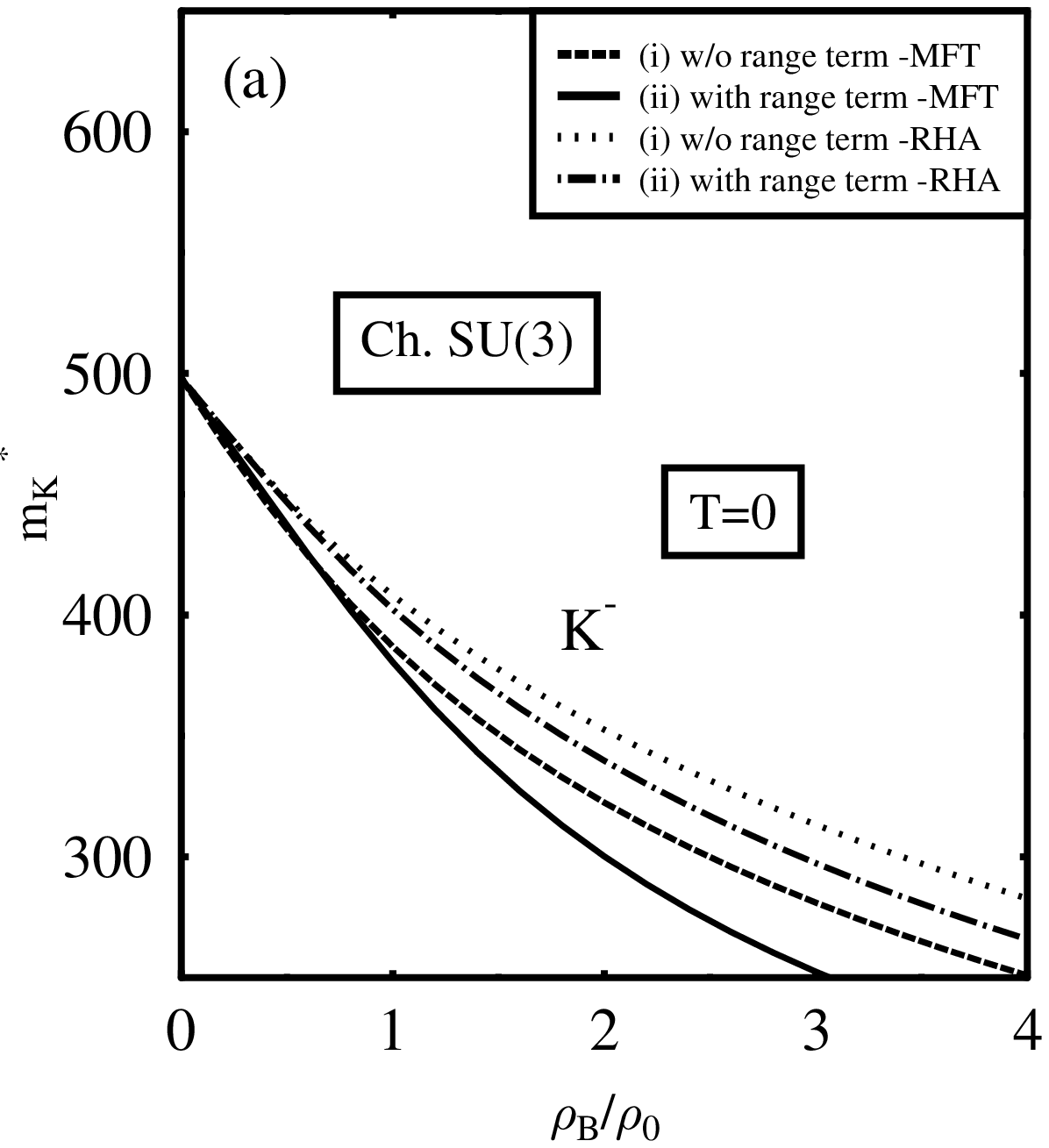}}
\parbox[b]{8cm}{
\includegraphics[width=9.2cm,height=9cm]{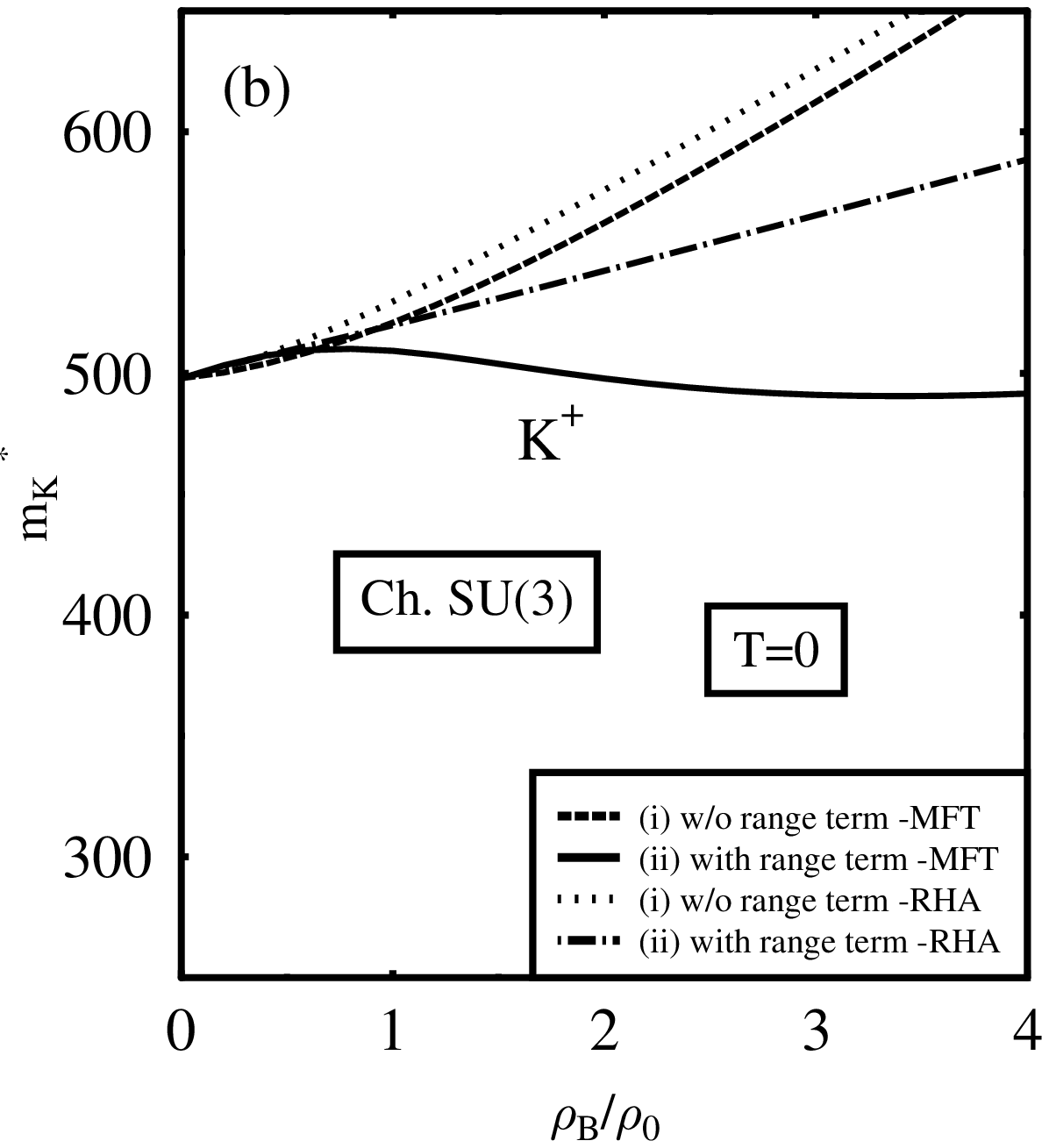}}}
\caption{
\label{mk0}
Masses of $K^\pm$ mesons due to the interactions in the
chiral SU(3) model at $T=0$, (i) without and (ii) with the contribution
from the range term.
}
\end{center}
\end{figure}

In Fig. 3 the masses of the $K$-mesons are plotted for $T=0$
in the present chiral model. We first consider the situation
when the Weinberg-Tomozawa term is supplemented by the
scalar and vector meson exchange interactions \cite{kmeson}.
The other case corresponds to the inclusion
of the range terms in (\ref{lagd}).
Both the $K$ meson mass as well as the $\bar K$ mass drop
at large densities when the range term is included.

\begin{figure}
\begin{center}
\centerline{\parbox[b]{8cm}{
\includegraphics[width=9.2cm,height=9cm]{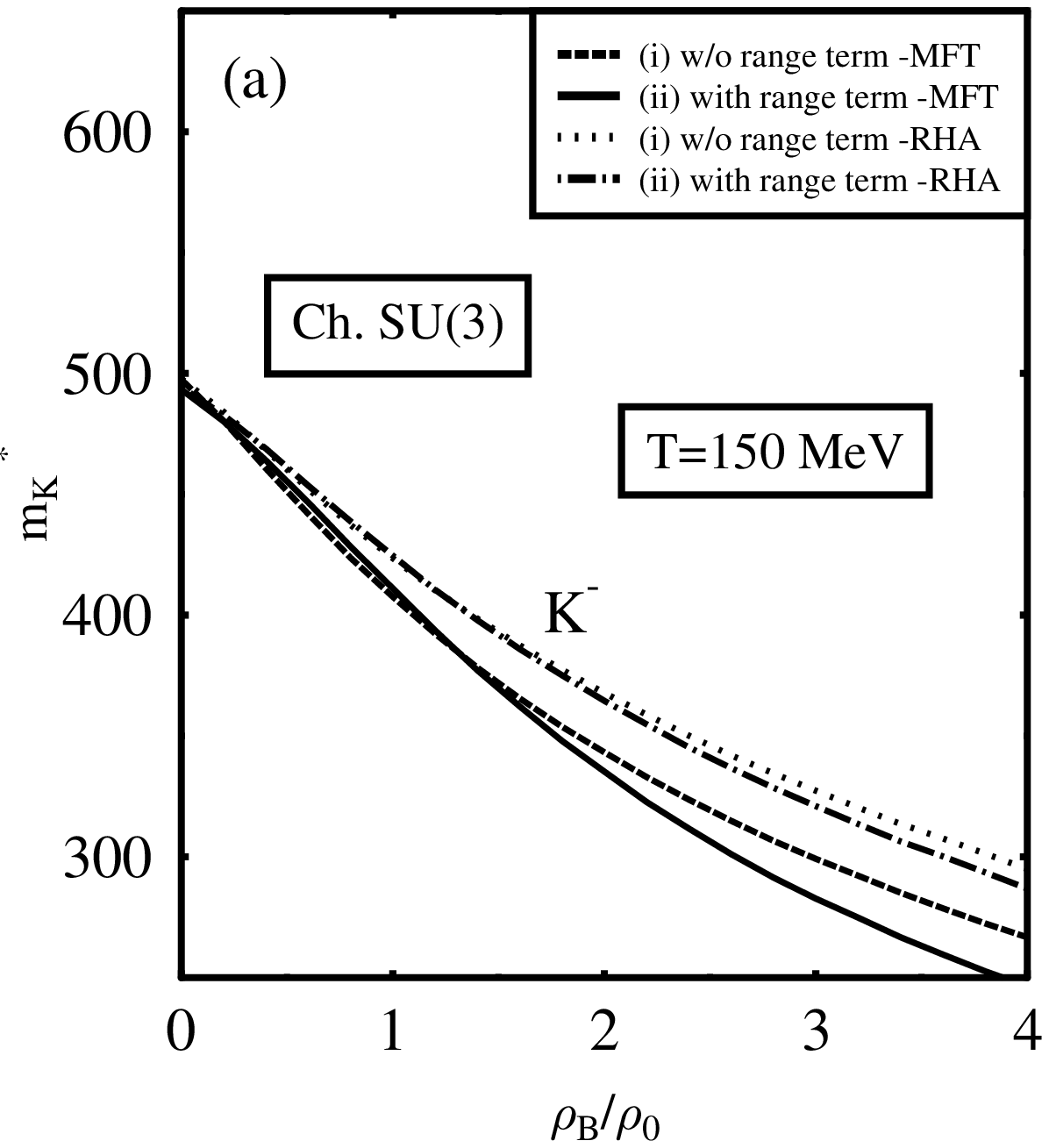}}
\parbox[b]{8cm}{
\includegraphics[width=9.2cm,height=9cm]{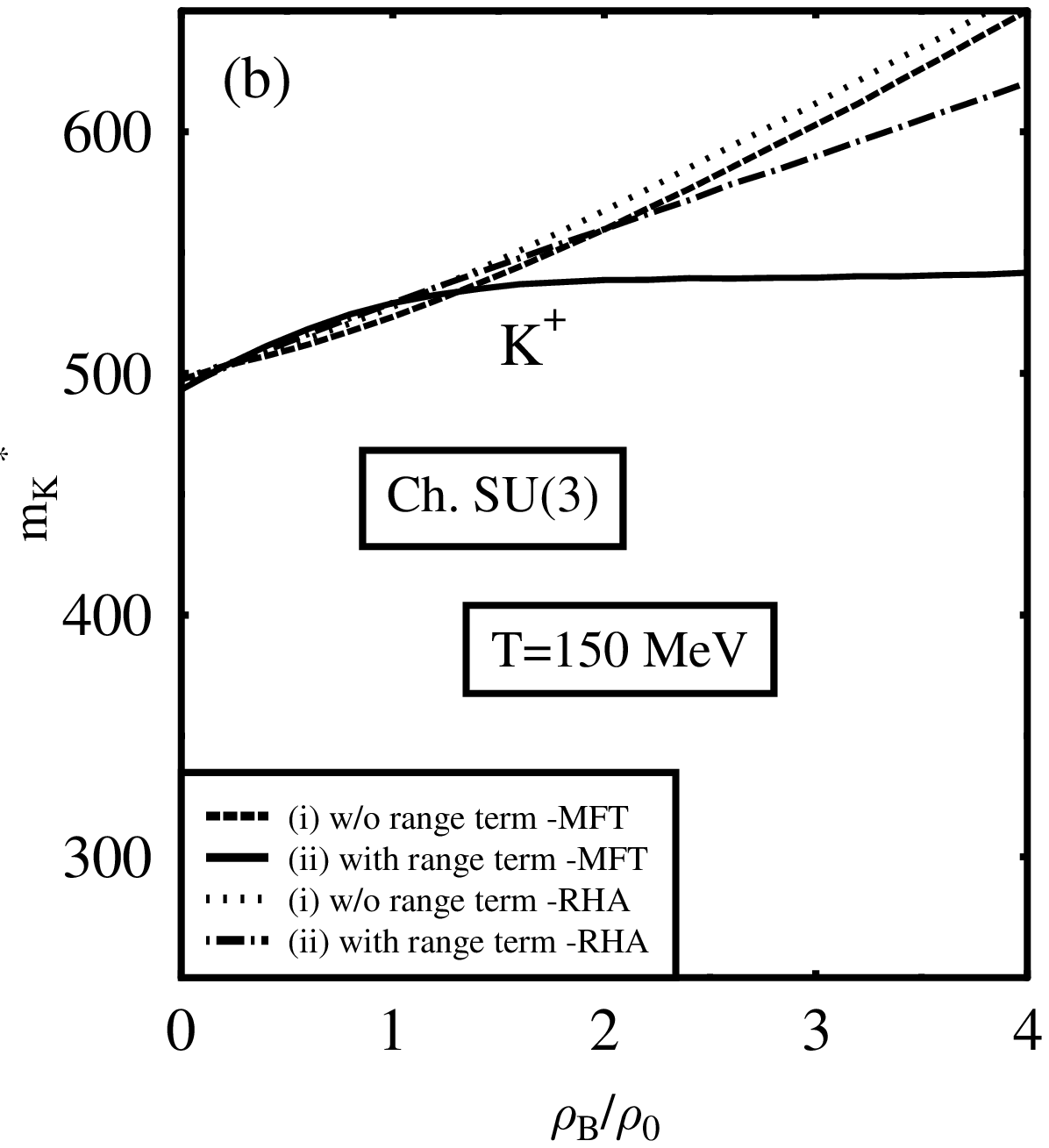}}}
\caption{
\label{mk150}
Masses of $K^\pm$ mesons in the chiral SU(3) model for $T=150$~MeV,
(i) without and (ii) with the range term contribution.
}
\end{center}
\end{figure}

In Fig. \ref{mk150}, the masses are plotted at temperature of 150 MeV.
The drop of the kaon masses are smaller as compared to the zero
temperature case. This is due to the fact that the nucleon mass
increases with temperature at finite densities in the chiral model used
here \cite{liko,kristof1}. Such a behaviour of the nucleon mass with
temperature was also observed earlier within the Walecka model by Ko
and Li \cite {liko} in a mean field calculation. The subtle behaviour
of the baryon self energy can be understood as follows. The scalar self
energy (\ref{denpart}) in the mean field approximation, increases due
to the thermal distribution functions at finite temperatures, whereas
at higher temperatures there are also contributions from higher momenta
which lead to lower values of the self energy. These competing effects
give rise to the observed increase of the effective baryon masses with
temperature at finite densities. This change in the nucleon mass with
temperature at finite density is also reflected in the vector meson
($\omega$, $\rho$ and $\phi$) masses in the medium \cite{kristof1}.
However at zero density, due to effects arising only from the thermal
distribution functions, the masses are seen to drop continuously with
temperature.

\begin{figure}
\begin{center}
\centerline{\parbox[b]{8cm}{
\includegraphics[width=9.2cm,height=9cm]{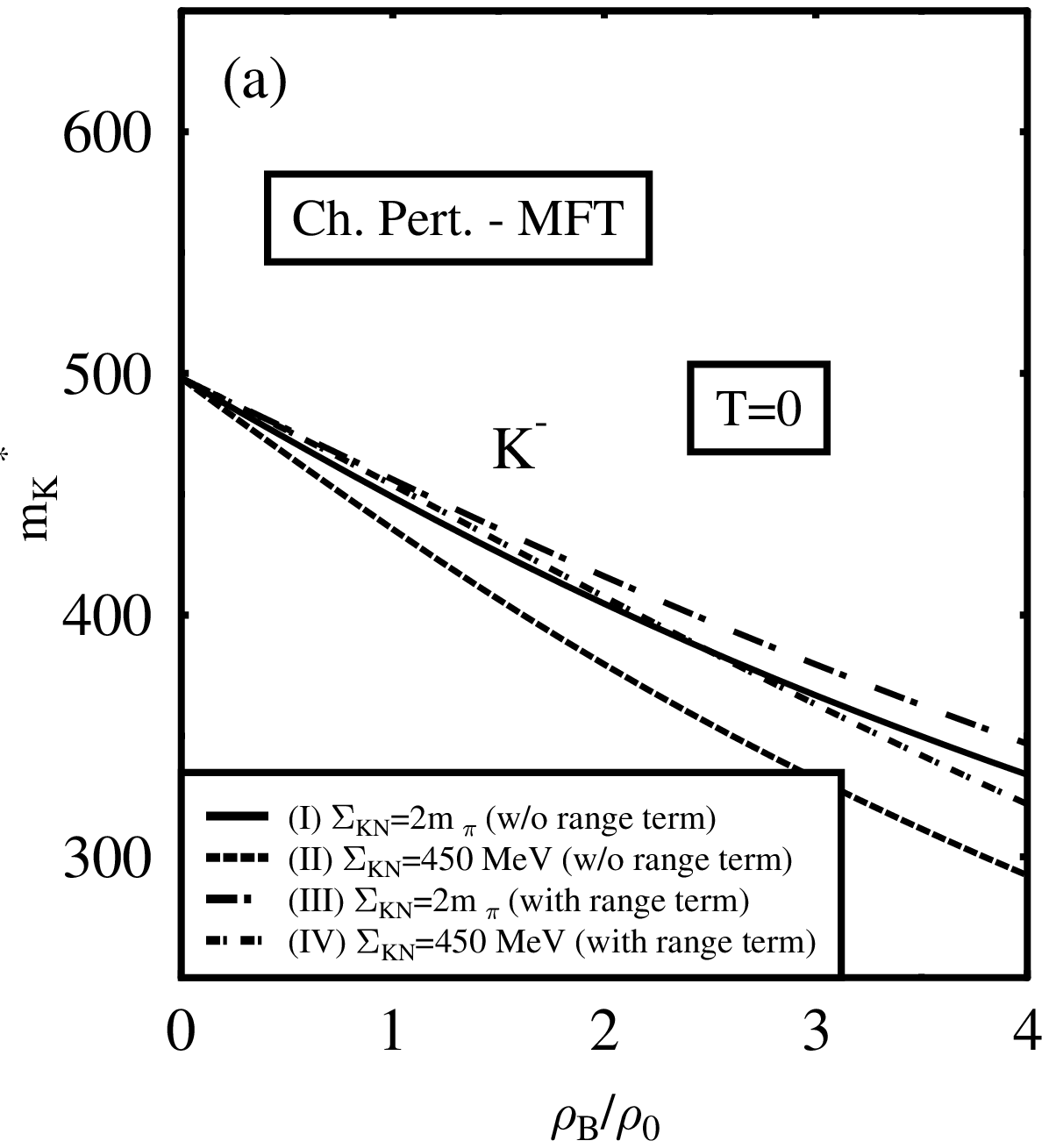}}
\parbox[b]{8cm}{
\includegraphics[width=9.2cm,height=9cm]{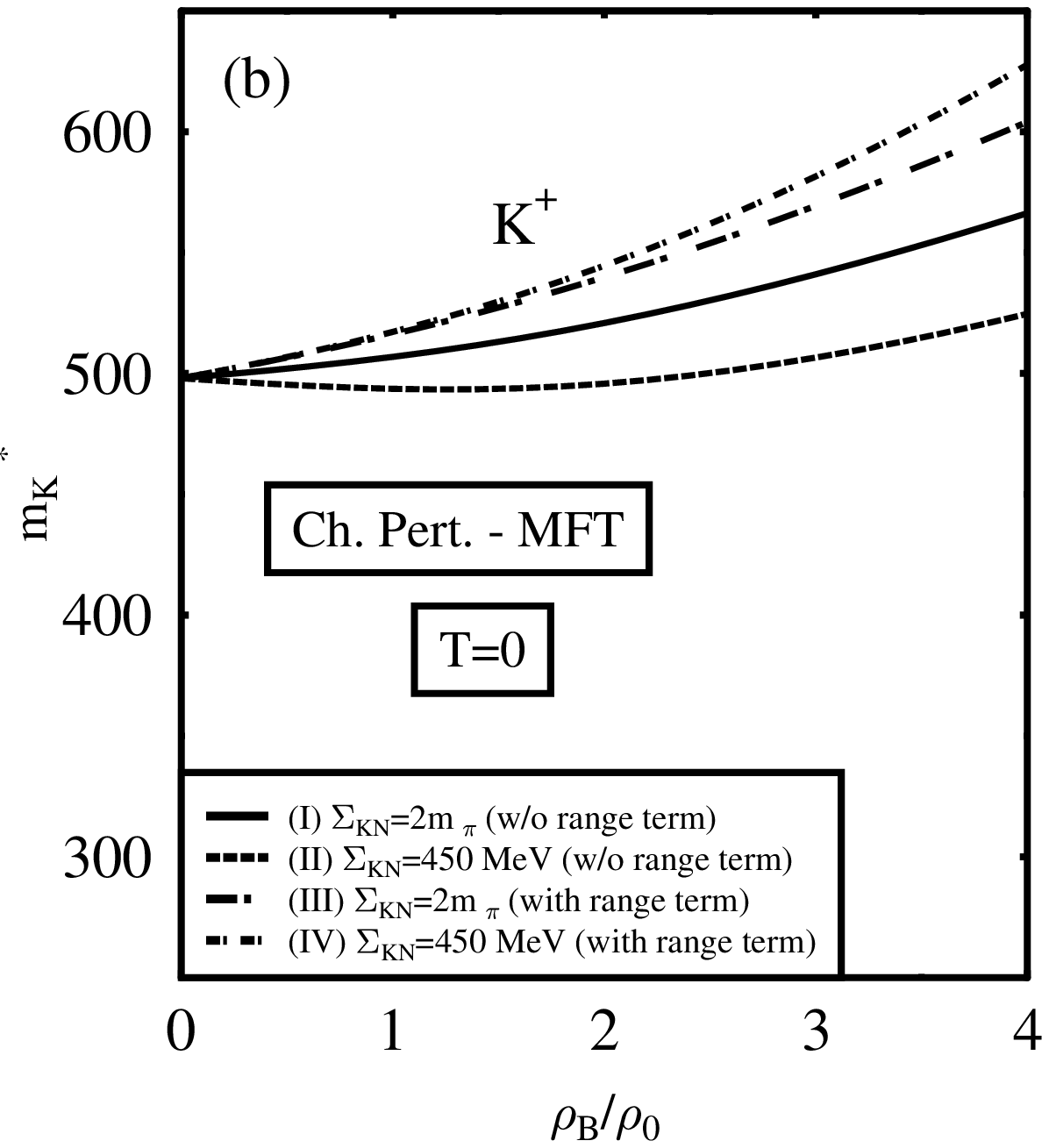}}}
\caption{
\label{mkt0mftcpt}
Masses of $K^\pm$ mesons at $T=0$ in the mean field approximation
in chiral perturbation theory, without and with
the contribution from range term.}
\end{center}
\end{figure}

\begin{figure}
\begin{center}
\centerline{\parbox[b]{8cm}{
\includegraphics[width=9.2cm,height=9cm]{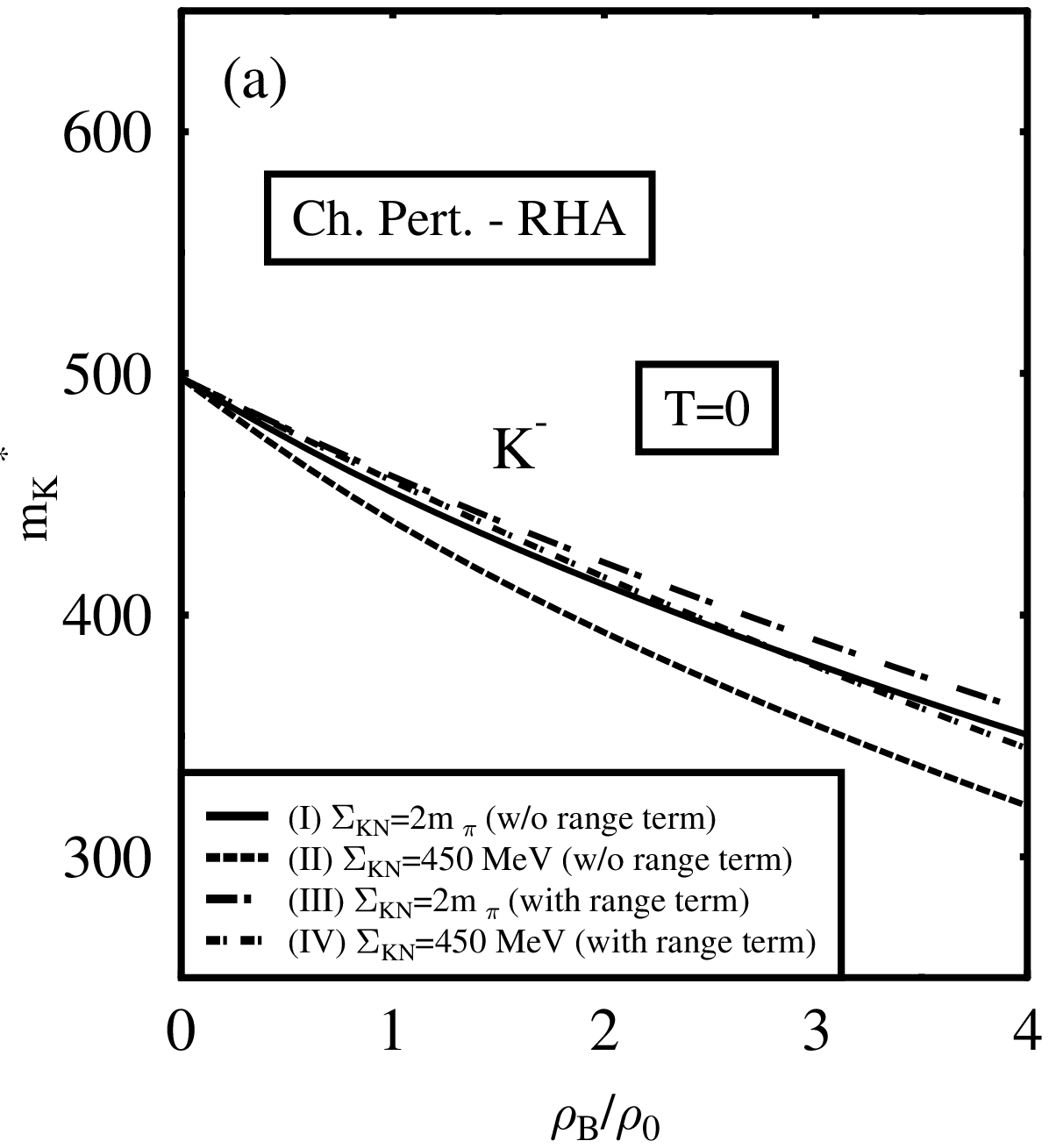}}
\parbox[b]{8cm}{
\includegraphics[width=9.2cm,height=9cm]{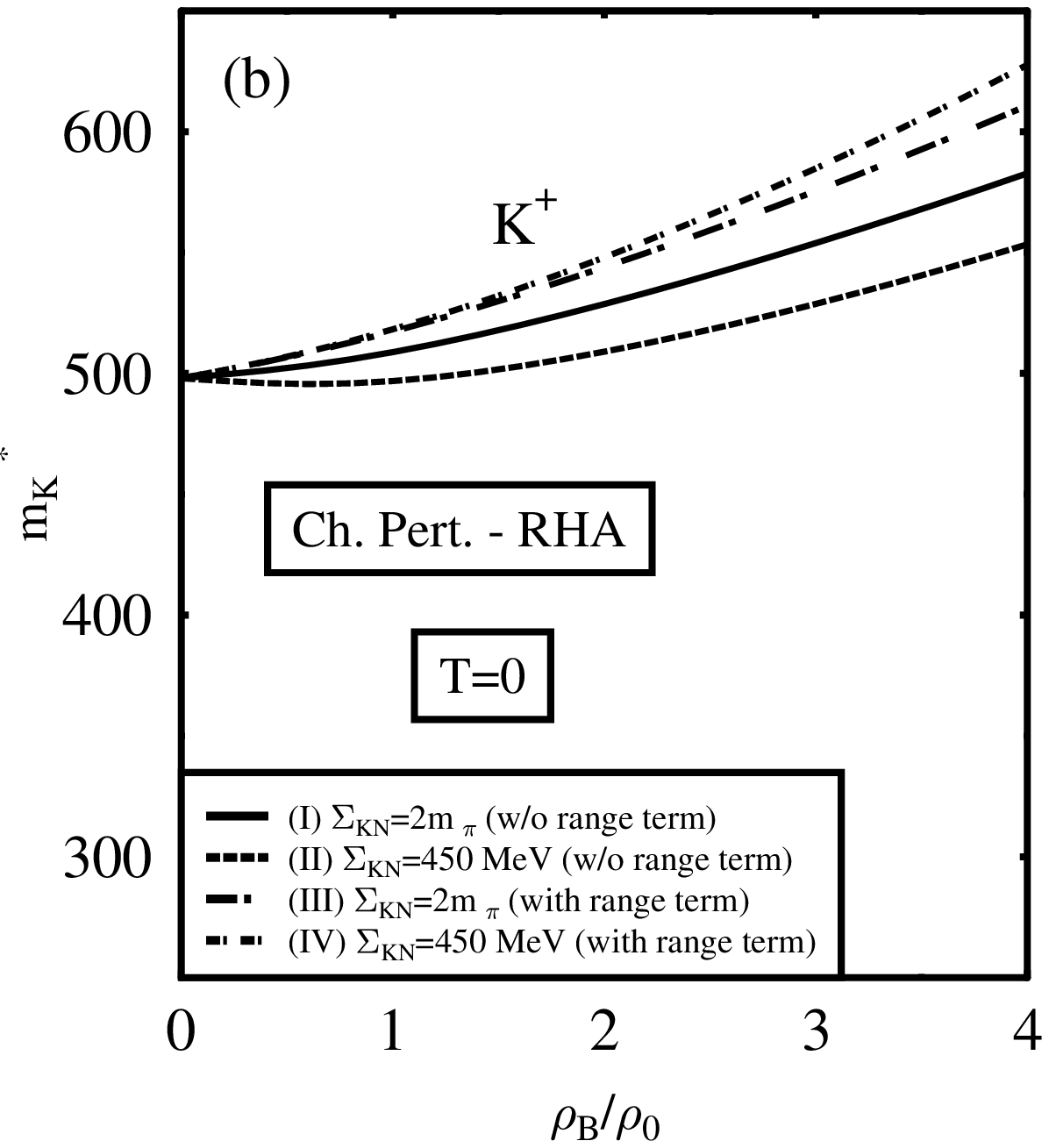}}}
\caption{
\label{mkt0rhacpt}
Same as in \ref{mkt0mftcpt}, but in the relativistic Hartree approximation.}
\end{center}
\end{figure}

We compare the results obtained in the chiral effective model to those
of the chiral perturbation theory (see ref. \cite{juergen,thorsson}).
The corresponding kaon masses are plotted in Figs. \ref{mkt0mftcpt} and
\ref{mkt0rhacpt} at zero temperature for the mean field as well as for
the relativistic Hartree approximation.  The $K^\pm$ masses are plotted
for different cases: (I) and (II) correspond to
$\Sigma_{KN}= 2m_\pi$ and $\Sigma_{KN}= 450$ MeV
respectively, without the range term, while the cases (III) and (IV)
are with the range term $(\partial _\mu K^+) (\partial ^\mu K^-)$, with
the parameter $\tilde D$ fitted to scattering length, so as to fulfill
the low density theorem \cite{thorsson}.  The case (II) shows a
stronger drop of the $K^-$ mass in the medium as compared to the case
(I) due to the larger attractive sigma term.  For $K^+$ however there
are cancelling  effects from the sigma term and the Weinberg-Tomozawa
interactions leading to only moderate mass modification.  The inclusion
of the repulsive range term in (III) and (IV) gives rise to a smaller
drop of the $K^-$ mass as compared to (I) and (II) respectively, where
this term is absent. For $K^+$, the term gives higher values for the
in-medium mass at large densities, as expected .  The relativistic
Hartree approximation shows again smaller mass modification as compared
to the mean field case. The K$^+$-meson mass shows a strong drop at large
density in the chiral effective model as compared to the other approaches.
The range term proportional to $d_1$ in (\ref{lagd}) has to overcome the
repulsive $\omega$- exchange term, to be compatible with the KN
scattering data, which is absent in chiral perturbation theory.  This
effect leads to a range term which is attractive contrary to the
situation in chiral perturbation theory where the range term is
repulsive. As a result, the effective chiral model gives stronger
modifications for the $K$-meson masses as compared to chiral
perturbation theory, especially at large density.

We note, that the somewhat shallow attractive potential for $K^-$ of
-30 to - 80 MeV has been extracted also from coupled-channel
calculations \cite{Lutz98,Lutz021,Lutz02,Oset00,Effenber00,lauran,Laura}
when including effects from dressing the $K^-$ propagator
selfconsistently.

Let us compare the behaviour of $K$ meson masses with the mass
modification of the D-meson \cite{dmeson} in a medium. The masses of
the $K^-$ as well as of the $D^+$ drop in the medium. For kaons, the
vector interaction in the chiral perturbation theory is the leading
contribution giving rise to a drop (increase) of the mass of $K^-$
($K^+$).  The subleading contributions arise from the sigma and range
terms with their coefficients as fitted to the KN scattering data
\cite{barnes}.  Fixing the charm sigma term by a generalized
GOR-relation the $D^-$ mass also increases in the medium \cite{dmeson}
similar to the behaviour of $K^+$. However, choosing the value for
$\Sigma_{DN}$ as calculated in the chiral effective model generalized
to SU(4), the mass of $D^-$ drops in the medium \cite{dmeson}.  In the
chiral effective model, the scalar exchange term as well as the range
term (which turns attractive for densities above $0.4 \rho_0$) lead to
the drop of both $D^+$ and $D^-$ masses in the medium.  We note here
that a similar behaviour is also obtained for the $K^+$.  Firstly, it
increases up to around a density of $0.8 \rho_0$ and then drops due to
the range term becoming attractive at higher densitites. However,
though the qualitative features remain the same, the medium
modification for $K^+$ is much less pronounced as compared to that of
the $D^-$ in the medium.  The medium modifications for the kaons and
D-mesons in either model are obtained in consistency with the low
energy KN scattering data.  The density modifications of the
$K(D)$-meson masses are seen to be large whereas the mass modifications
are seen to be rather insensitive to temperature.

\section{$K^\pm$ production at SIS energies within a covariant
transport model}

Since the different models discussed in the previous Sections give
very different results for the $K^\pm$ properties in the nuclear
medium, it is of central importance to obtain further information
from the experimental studies on $K^\pm$ production in order to
support or reject part of the models. However, high density matter
can only be produced in relativistic nucleus-nucleus collisions,
where $K^\pm$ production and propagation happens to a large extent
out of kinetic and chemical equilibrium. One thus has to employ
non-equilibrium transport approaches to follow the dynamics of all
hadrons in phase space \cite{CaMo}.

\subsection{Description of the transport model}

Our study of heavy-ion collisions are based on the
hadron-string-dynamics (HSD) transport approach
\cite{Cass97,brat97,CB99}.  Though the HSD transport approach has been
developed for the off-shell dynamics including the propagation of
hadrons with dynamical spectral functions \cite{Cass99off} and also has
been used  in the context of $K,\bar{K}$ production and propagation in
nucleus-nucleus collisions \cite{laura03}, we here restrict to the
on-shell quasi-particle realization similar to Refs.
\cite{Cass97,brat97,CB99}. The main reason is that the full off-shell
calculations require the knowledge of the momentum, density and
temperature dependent spectral functions of $K,\bar{K}$ mesons as well
as the in-medium cross sections for all production and absorption
channels. The latter have to be calculated in a consistent way
incorporating again the same spectral functions.  Such information is
naturally provided by coupled-channel $G$-matrix calculations
\cite{laura03}.  However, the models presented in Sections II-V are not
suited for such purpose since they are formulated on the mean-field
level, only.  We note in passing that the differences between the
present on-shell and the previous off-shell \cite{laura03} versions of
transport for $\bar{K}$ spectra in the SIS energy regime are less than
30\% for the systems to be investigated below if similar antikaon
potentials are employed \cite{Cass_priv}.

In Refs. \cite{Cass97,brat97,CB99} the transition amplitudes for
$\bar{K} N$ channels below the threshold of $\approx$ 1.432 GeV have
been extrapolated from the vacuum amplitudes in an 'ad hoc' fashion,
which differ sizeable  from more recent microscopic coupled-channel
calculations \cite{Lutz98,Lutz021,lauran,Lutz02}. To reduce this
ambiguity we have adopted the results from the $G$-matrix calculations
of Ref.  \cite{lauran}, which have been also incorporated in the
off-shell calculations \cite{laura03} and extend far below the 'free'
threshold. The latter $G$-matrix calculations have been performed at a
fixed temperature $T$ = 70 MeV, which corresponds to an average
temperature of the 'fireballs' produced in nucleus-nucleus collisions
at SIS energies.  We recall that variations in the temperature from 50
- 100 MeV do not sensibly affect the quasi-particle properties in the
medium according to the studies in Ref. \cite{lauran}.

Actual cross sections in our present approach are determined as a
function of the invariant energy squared $s$ as \cite{laura03}
\begin{equation}
\label{cross} \sigma_{1+2 \rightarrow 3+4}(s) = (2 \pi)^5
\frac{E_1 E_2 E_3 E_4}{s} \frac{p'}{p} \ \int d\cos( \theta) \
\frac{1}{(2s_1+1)(2s_2+1)} \sum_i \sum_\alpha \ G^\dagger G,
\end{equation}
where $p$ and $p'$ denote the center-of-mass momentum of the particles
in the initial and final state, respectively, and $E_j$ stand for the
particle energies. The sums over $i$ and
$\alpha$ indicate the summation over initial and final spins, while
$s_1, s_2$ are the spins of the particles in the entrance channel.
Apart from the kinematical factors, the transition rates are determined
by the angle integrated average transition probabilities
\begin{equation}
\label{crossp} P_{1+2 \rightarrow 3+4}(s) =  \int d\cos(\theta) \
\frac{1}{(2s_1+1)(2s_2+1)} \sum_i \sum_\alpha \ G^\dagger G
\end{equation}
which -- as mentioned above -- are uniquely determined by the $G$-matrix
elements evaluated for
finite density $\rho$, temperature $T$ and relative momentum $p_{\bar
  K}$ with respect to the nuclear matter rest frame.
The transition probabilities of Eq.~(\ref{crossp})  have been
displayed in the r.h.s. of Figs. 5-8 of Ref. \cite{laura03}
for the reactions $K^- p \rightarrow K^- p$, $K^- p
\rightarrow \Sigma^0 \pi^0$, $K^- p \rightarrow \Lambda \pi^0$ and
$\Lambda \pi^0 \rightarrow \Lambda \pi^0$ as a function of density and
invariant energy, respectively. The latter have been parametrized
by the authors of Ref. \cite{laura03} and are available to the
public \cite{Wolfgang}.

In this context it is important to point out that the backward channels
$K^- p \leftarrow \Sigma^0 \pi^0$, $K^- p \leftarrow \Lambda \pi^0$
etc.  are entirely determined by detailed balance, which is strictly
fulfilled in the HSD transport approach using Eq. (\ref{cross}).

In principle, the real parts of the $\bar{K}$ self energies are also
fully determined by the $G$-matrix calculations.  However,
we here a adopt a {\em hybrid model} that keeps the in-medium transitions
probabilities (\ref{crossp}) fixed and vary the real part of the
$\bar{K}$ self energies or antikaon potential according to the models
presented in Sections II-V. In this way one can study the explicit
effect of the kaon and antikaon potentials in the nuclear medium in a
more transparent way without employing too rough approximations for the
transition probabilities involving antikaons. These transition
amplitudes are beyond the level of mean-field theory essentially
discussed in Sections II to V.

We stress that for the present study we employ the kaon production
cross sections for $N \Delta$ and $\Delta \Delta$ channels from Ref.
\cite{Tsushima} instead of the previously used fixed isospin relations, i.e.
$\sigma_{N\Delta \rightarrow NKY} (\sqrt{s}) = 3/4 \sigma_{NN \rightarrow NKY} (\sqrt{s})$
and $\sigma_{\Delta\Delta \rightarrow NKY} (\sqrt{s}) = 1/2
\sigma_{NN \rightarrow NKY} (\sqrt{s})$.
The kaon yields in vacuum now are on average
enhanced by $\sim$ 30\% relative to the yields in Refs.
\cite{Cass97,brat97,CB99}.  This enhancement is a
consequence of the larger production cross section in the $N \Delta$
and $\Delta \Delta$ channels from Ref. \cite{Tsushima}
(as also used in Refs. \cite{Aichelin,Fuchs}).
Since these resonance induced production cross sections cannot be
measured in vacuum, the actual $K^+$ yield from A+A collisions
calculated with transport models might differ substantially
depending on the parametrizations involved.

\subsection{$K^\pm$ spectra from nucleus-nucleus collisions}

\begin{figure}[!]
\centerline{\psfig{file=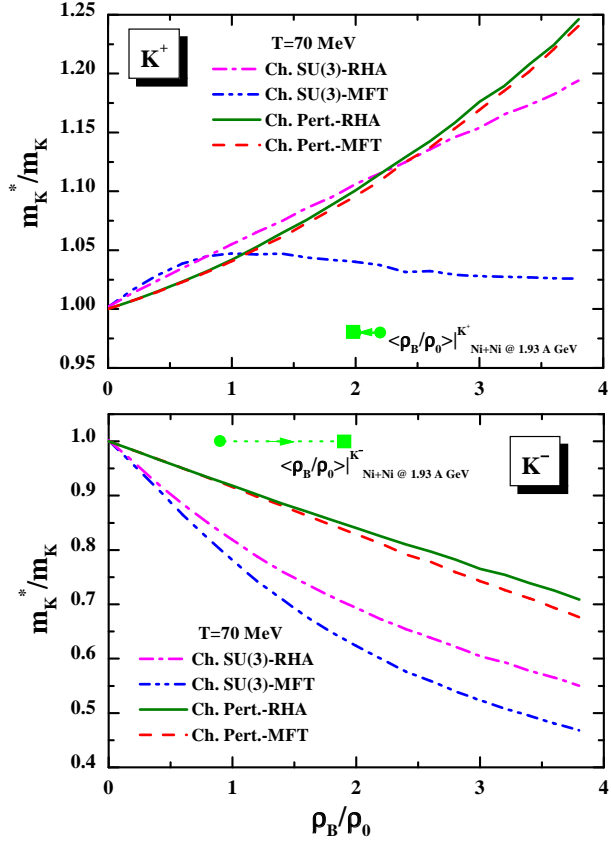,width=8cm}}
\caption{The ratio of the in-medium kaon (upper part) and antikaon
(lower part) masses to the vacuum masses ($m_K^*/m_K$) as a function of
baryon density in units of $\rho_0=0.16$ fm$^{-3}$ calculated in the
different models for the temperature $T=70$ MeV. The dot-dashed lines
correspond to the results within the chiral SU(3) model in the
relativistic Hartree approximation ('Ch. SU(3)-RHA'), the
dot-dot-dashed lines stand for the chiral SU(3) model in the mean-field
approximation ('Ch. SU(3)-MFT'), the solid lines indicate the
calculations within the chiral perturbation theory in relativistic
Hartree approximation ('Ch. Pert.-RHA'), whereas the dashed lines show
the results for the chiral perturbation theory in mean-field
approximation ('Ch. Pert.-MFT').
The full symbols connected by arrows indicate the averaged freeze-out
baryon density in units of $\rho_0$ of $K^+$ (upper part) and $K^-$
mesons (lower part)  for free (dots) and in-medium scenarios (squares)
for central ($b=1$ fm) Ni+Ni collisions at 1.93 A$\cdot$GeV.}
\label{mass-k}
\end{figure}

For reasons of transparency we show in Fig. \ref{mass-k} the results
from Sections II to V for the kaon and antikaon masses in nuclear
matter at a temperature $T$= 70 MeV that are parametrized as a function
of the density $\rho_B$ and enter the transport calculation in the
production as well as propagation parts. In Fig. \ref{mass-k} the
dot-dashed lines correspond to the results within the chiral SU(3)
model in the relativistic Hartree approximation ('Ch. SU(3)-RHA'), the
dot-dot-dashed lines stand for the chiral SU(3) model in the mean-field
approximation ('Ch. SU(3)-MFT'), the solid lines indicate the
calculations within the chiral perturbation theory in relativistic
Hartree approximation ('Ch. Pert.-RHA'), whereas the dashed lines show
the results for the chiral perturbation theory in mean-field
approximation ('Ch. Pert.-MFT'). Whereas the results of the different
models roughly coincide for kaons (upper part) - except for the limit
'Ch. SU(3)-MFT', the modifications of the $K^-$ masses (lower part)
differ more drastically. Here the 'Ch.  SU(3)-MFT' limit gives the
lowest masses, followed by 'Ch.  SU(3)-RHA'. The results from the two
limits of chiral perturbation theory here provide the lowest mass
modifications.

The full symbols connected by arrows indicate the averaged freeze-out
baryon density in units of $\rho_0$ of $K^+$ (upper part) and $K^-$
mesons (lower part)  for free (dots) and in-medium scenarios (squares)
for central ($b=1$ fm) Ni+Ni collisions at 1.93 A$\cdot$GeV.  Since
$K^+$ mesons are basically produced from primary collisions and suffer
less from rescattering, they see a higher baryon density $\sim 2
\rho_0$. The $K^-$ mesons are freezing-out later and dominantly stem
from the pion-hyperon interactions. Since the difference in the $K^-N$
and $\pi \Lambda$ thresholds for the non-modified $K^-$ masses is about
320 MeV, the $K^-$ mesons can be produced only by energetic pions or
hyperons.  Consequently, the density at the production point is only
$\sim \rho_0$, i.e. lower than the initial baryon density achieved,
e.g., in Ni+Ni collisions at 1.93 A$\cdot$GeV. The in-medium shift of
$K^-$ masses  reduces the thresholds for $K^-$ production, such that
$K^-$ mesons are created earlier (and more frequent) by low momentum
particles at high baryon density $\sim 1.8 \rho_0$ (depending on
the strength of the attractive potential).
Moreover, the in-medium absorption cross section of antikaons according
to the transition probabilities employed here, is lower than the free
absorption cross section.

Thus, $K^+, K^-$ mesons produced in heavy-ion collisions probe the
density regime $\sim 1\div 2 \rho_0$, where the different models
deviate substantially, so one can hope to check the reliability of the
models by comparing to the experimental data for A+A reactions.

\begin{figure}[!]
\phantom{a}\hspace*{-4cm}
\psfig{file=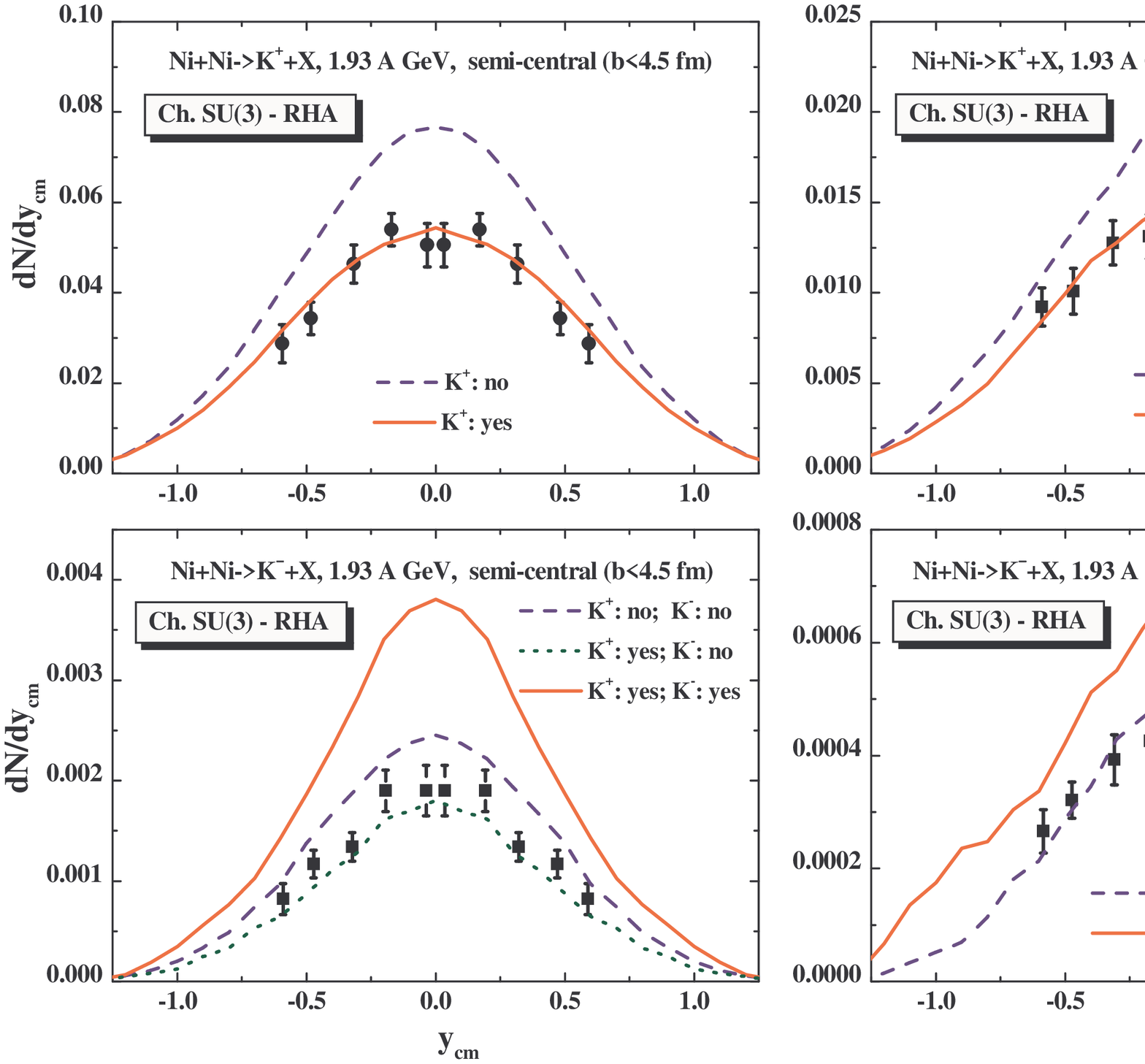,width=11cm}
\caption{The rapidity spectra of $K^+$ (upper part) and $K^-$ mesons
(lower part) for Ni+Ni at 1.93 A$\cdot$GeV for semi-central
($b\leq 4.5$~fm) (l.h.s.) and non-central ($b=4.5\div 7.5$~fm) (r.h.s)
collisions in comparison to the KaoS data from Ref.
\protect\cite{kaosnew}.  The dashed lines correspond
to the 'free' calculations, i.e. discarding  $K^+$ and $K^-$ potentials,
the dotted line (lower left plot) shows the
calculations with a $K^+$ potential according to the chiral SU(3)-RHA
model, the solid lines correspond to the results for the
$K^+$ and $K^-$ potentials from the chiral SU(3)-RHA model.}
\label{yNi19_RHA}
\end{figure}

We start with differential spectra for $K^+$ and $K^-$ mesons from
Ni+Ni reactions at 1.93 A$\cdot$GeV for semi-central
($b\leq 4.5$~fm) (l.h.s.) and non-central ($b=4.5\div 7.5$~fm) (r.h.s)
collisions in comparison to the KaoS data from Ref.
\protect\cite{kaosnew} (Fig. \ref{yNi19_RHA}). The dashed lines correspond
to the 'free' calculations, i.e. discarding  $K^+$ and $K^-$ potentials,
the dotted line (lower left plot) shows the
calculations with a $K^+$ potential according to the chiral SU(3)-RHA
model, the solid lines correspond to the results for the
$K^+$ and $K^-$ potentials from the chiral SU(3)-RHA model. It is
seen that the $K^+$ rapidity distributions are overestimated in
comparison to the data when discarding a kaon potential. However,
when including the repulsive $K^+$ potential from the chiral SU(3)-RHA
model a very satisfactory agreement is achieved for semi-central
and non-central collisions.

The $K^-$ rapidity distributions are
slightly overestimated when using only 'free' kaon and antikaon
masses. This limit, however, is unphysical since the $K^+$ spectra
require a repulsive potential as demonstrated in the upper part
of the figure. Now, when including the repulsive $K^+$ potential,
the $K^-$ rapidity distribution is slightly underestimated due to
the lower amount of $K^-$ production by the hyperon+pion
production channel. Note, that the hyperon abundance is strongly
correlated with the abundance of kaons due to strangeness
conservation. Incorporating  the very strong $K^-$ potential from
the chiral SU(3)-RHA model, however, the antikaon
yields are severely overestimated by a factor $\sim$ 1.6 to 1.8.
This result indicates that the attraction for the $K^-$ in the
chiral SU(3)-RHA model is too strong.

\begin{figure}[!]
\phantom{a}\hspace*{-4cm}
\psfig{file=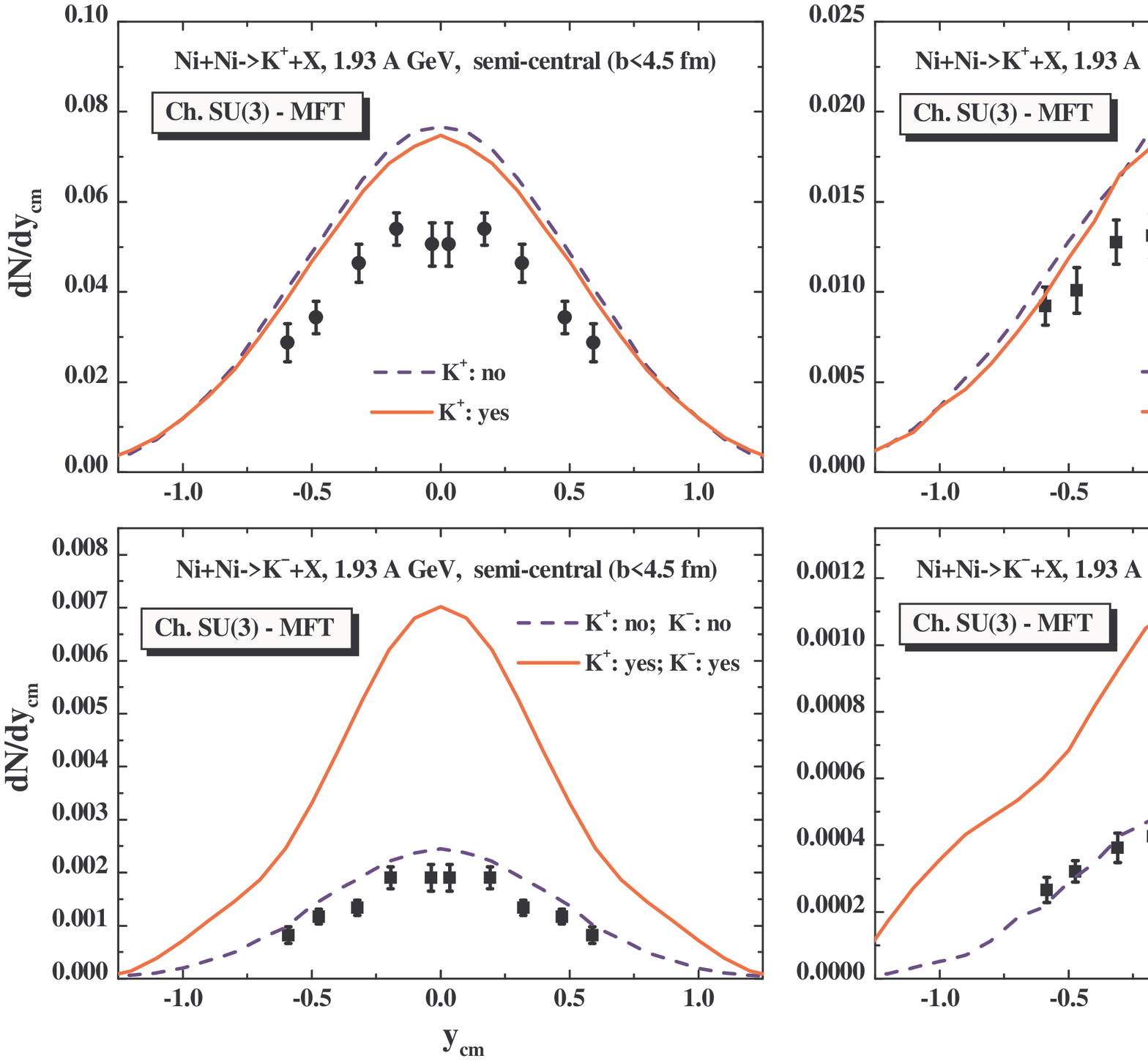,width=11cm}
\caption{The rapidity spectra of $K^+$ (upper part) and $K^-$ mesons
(lower part) for  Ni+Ni at 1.93 A$\cdot$GeV for semi-central ($b\leq
4.5$~fm) (l.h.s.) and non-central ($b=4.5\div 7.5$~fm) (r.h.s.)
collisions in comparison to the KaoS data from Ref.
\protect\cite{kaosnew}.  The dashed lines  correspond
to the 'free' calculations, while the solid lines
show the results for the $K^+$ and $K^-$ potentials from the chiral
SU(3)-MFT model.}
\label{yNi19_MFT}
\end{figure}

In Fig. \ref{yNi19_MFT} we show the same comparison as in Fig.
\ref{yNi19_RHA} for the  chiral SU(3)-MFT model. In this case the kaon
potential is very low and practically does not give a
sufficient reduction of the kaon rapidity distribution relative to the
'free' case. Consequently, also the hyperon abundance is overestimated
in this model, which gives too many antikaons (by the $\pi$+hyperon
channels) already without including any $\bar{K}$ potential. On the
other hand,  the chiral SU(3)-MFT model leads to very strong $K^-$
potentials which in turn give $K^-$ yields in nucleus-nucleus
collisions that are larger than the data by up to a factor of 3.  We
conclude that this model is essentially ruled out by the data due to i)
a lacking repulsion in the kaon channel and ii) too much attraction in
the antikaon channel.

\begin{figure}[!]
\phantom{a}\hspace*{-4cm}
\psfig{file=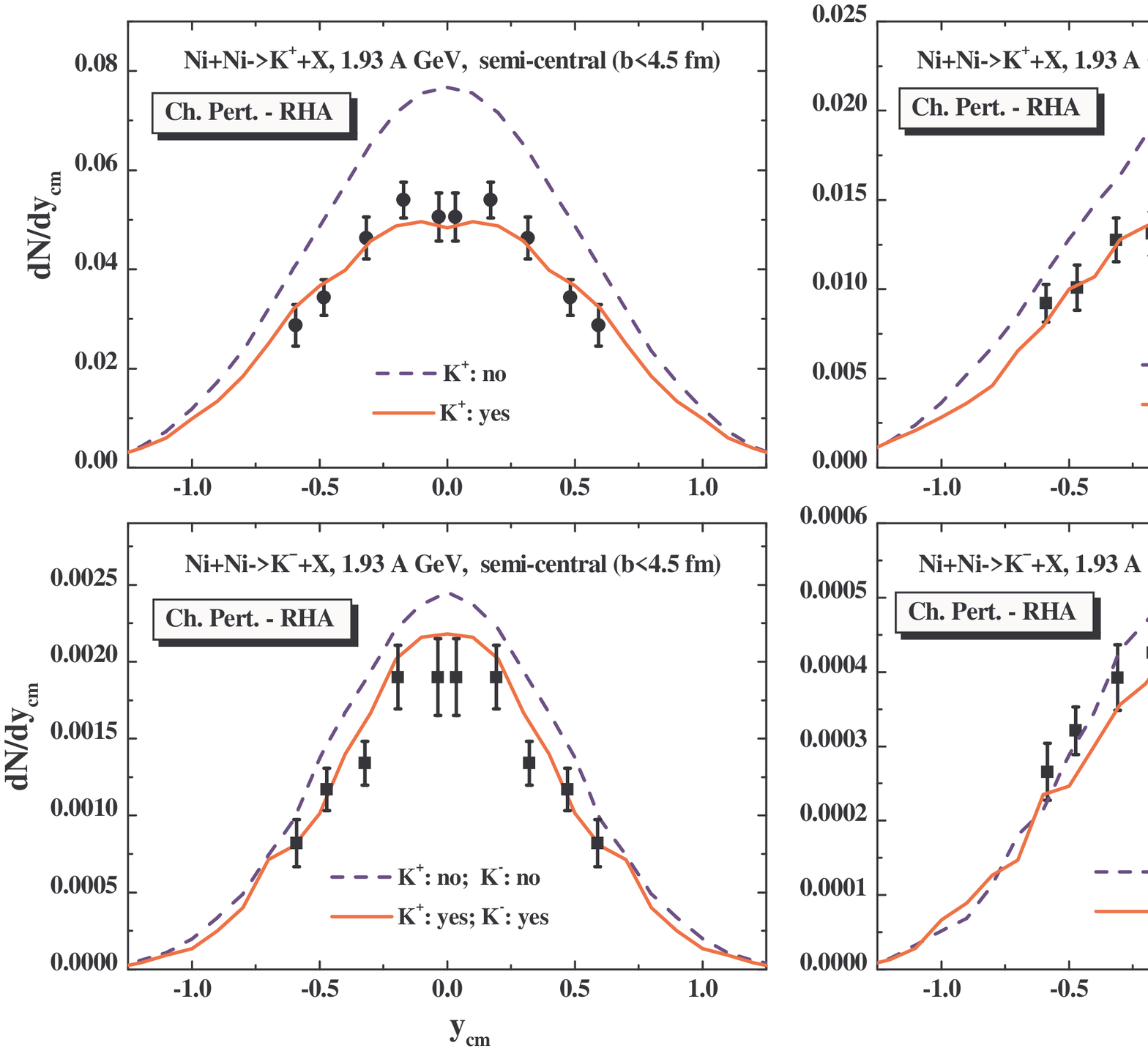,width=11cm}
\caption{The rapidity spectra of $K^+$ (upper part) and $K^-$ mesons
(lower part) for Ni+Ni at 1.93 A$\cdot$GeV for semi-central
($b\leq 4.5$~fm) (l.h.s.) and non-central ($b=4.5\div 7.5$~fm) (r.h.s.)
collisions in comparison to the KaoS data from Ref.
\protect\cite{kaosnew}.  The dashed lines correspond
to the 'free' calculations, while the solid lines
show the results for the $K^+$ and $K^-$ potentials from the chiral
perturbation theory in relativistic Hartree approximation ('Ch.
Pert.-RHA').}
\label{yNi19-CP}
\end{figure}

In Fig. \ref{yNi19-CP} we continue with the same comparison as in Fig.
\ref{yNi19_RHA} for the  chiral perturbation theory in relativistic
Hartree approximation ('Ch.  Pert.-RHA'). In this case the kaon
potential potential is moderately repulsive  and gives a sufficient
reduction of the kaon rapidity distribution relative to the 'free' case
almost perfectly in line with the data. When including now additionally
the in-medium modifications the $K^-$ rapidity distributions are in a
very good agreement with the data. We thus find that the chiral
perturbation theory gives the proper repulsion in the kaon as well as
attraction in the antikaon channel.

\begin{figure}[!]
\centerline{\psfig{file=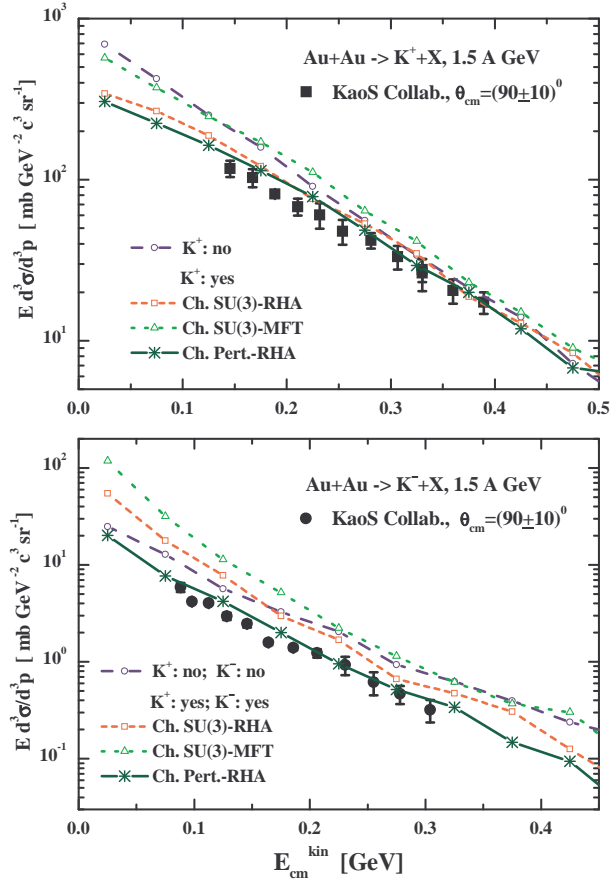,width=8cm}}
\caption{The differential inclusive $K^+$ (upper part) and $K^-$
spectra (lower part) for Au+Au at 1.48 A$\cdot$GeV and $\theta_{cm} =
(90 \pm 10)^o$ in comparison to the KaoS data from Ref.
\protect\cite{Forster02}.  The dashed lines with open circles correspond to
the 'free' calculations, the dashed lines with open squares stand for
the results with the $K^+$ and $K^-$ potentials from the chiral
SU(3)-RHA model, the dotted lines with open triangles indicate the
calculations within the chiral SU(3)-MFT model and the solid lines with
stars show the results with the $K^+$ and $K^-$ potentials from chiral
perturbation theory in relativistic Hartree approximation.}
\label{S3_Au15}
\end{figure}

Though the Ni+Ni system at 1.93 A$\cdot$GeV already provides
clear hints for the proper size of $K^\pm$ self energies at
finite baryon density, it is important to check these indications
by independent observables in a different system, too. In this
respect we show in Fig. \ref{S3_Au15} the differential
inclusive $K^+$ (upper part) and $K^-$
spectra (lower part) for Au+Au at 1.48 A$\cdot$GeV and $\theta_{cm} =
(90 \pm 10)^o$ in comparison to the KaoS data from Ref.
\protect\cite{Forster02}. The dashed lines with open circles correspond to
the 'free' calculations, the dashed lines with open squares stand for
the results with the $K^+$ and $K^-$ potentials from the chiral
SU(3)-RHA model, the dotted lines with open triangles indicate the
calculations within the chiral SU(3)-MFT model and the solid lines with
stars show the results with the $K^+$ and $K^-$ potentials from chiral
perturbation theory in relativistic Hartree approximation.
We find that without any $K^+$ and $K^-$ potentials the $K^+$
spectra are overestimated and show a slope parameter which is too
low in comparison with experiment. Due to strangeness conservation
also the hyperons are too frequent in this limit, which leads to
an slight overestimation of the antikaon spectra. In this case the $K^-$
slope comes out too high.

As discussed in the context of Fig. \ref{yNi19_MFT} the chiral
SU(3)-MFT model yields only a small change of the $K^+$ spectra but
overestimates the $K^-$ spectra again by a factor of 3 -- 4. Thus we
obtain the same conclusion for the Au+Au system at 1.48 A$\cdot$GeV
when looking at the transverse kinetic energy spectra. In the chiral
SU(3)-RHA model the kaons come out reasonably well, but again the $K^-$
spectrum is severely overestimated due to the strong attraction in the
antikaon channel. Only the chiral perturbation theory (stars) provides
a good description of both spectra simultaneously also for the Au+Au
system at 1.48 A$\cdot$GeV, which gives a further indication for the
proper $K^\pm$ self energies given in this limit. We, furthermore,
stress that not only the magnitude of the differential $K^+, K^-$
spectra is described well in chiral perturbation theory, but also the
$K^+, K^-$ slopes (cf. Ref. \cite{laura03}).

\begin{figure}[!]
\centerline{\psfig{file=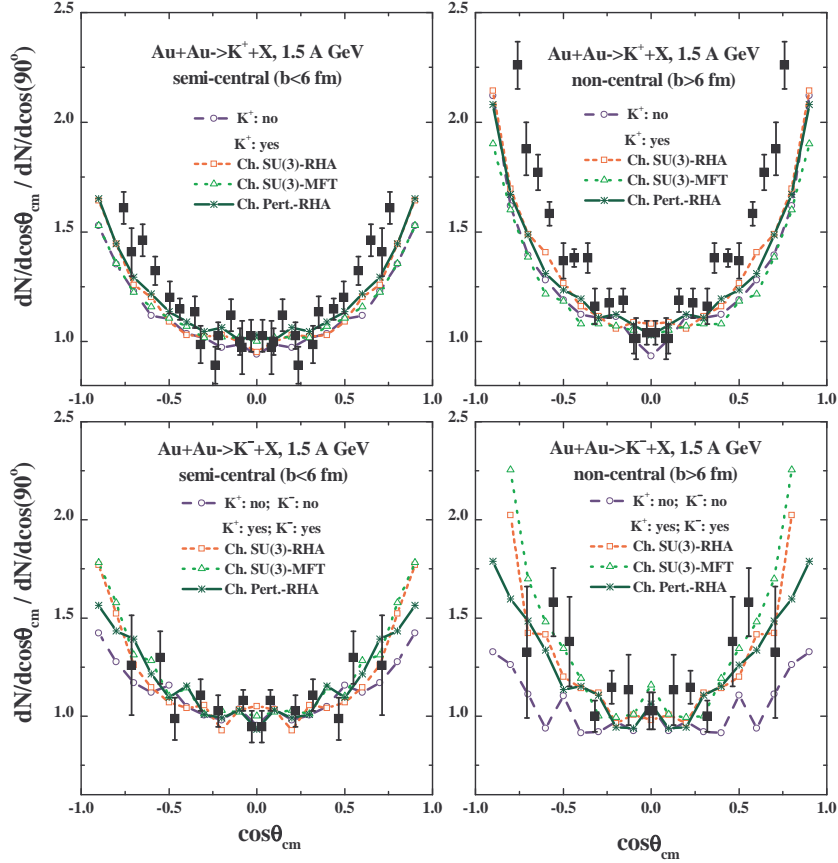,width=11cm}}
\caption{The $K^+$ (upper part) and $K^-$ angular distributions (lower
part) for semi-central (l.h.s.) and non-central (r.h.s.) Au+Au
collisions at 1.48 A$\cdot$GeV. All angular distributions are normalized
to unity for $\cos \theta_{cm}$ = 0.  The experimental data have been
taken from Ref. \protect\cite{Andreas}.  The assignment of the
individual lines is the same as in Fig. \protect\ref{S3_Au15}.}
\label{cos_au15}
\end{figure}

In Fig. \ref{cos_au15} we show additionally the $K^+$ (upper part) and
$K^-$ angular distributions (lower part) for semi-central (l.h.s.) and
non-central (r.h.s.) Au+Au collisions at 1.48 A$\cdot$GeV. We note that
all angular distributions have been normalized to unity for $\cos
\theta_{cm}$ = 0 as well as the experimental data from Ref.
\protect\cite{Andreas}.  The assignment of the individual lines from
the different models is the same as in Fig. \protect\ref{S3_Au15}. We
find that all models do not differ very much in the angular
distributions which are more sensitive to the elastic and inelastic
cross sections employed in the transport approach for kaons and
antikaons. Only for non-central collisions the $K^-$ angular
distribution comes out too flat in the 'free' case, i.e.  when no
$K^\pm$ potentials and 'free' transition rates for the scattering
processes are used in the transport model. We note that these angular
distributions are almost the same as in the full off-shell calculations
from Ref. \cite{laura03} for this system.

\subsection{Collective flow of kaons and antikaons}

Though the rapidity and transverse energy distributions for
$K^\pm$ mesons have given some preference for chiral perturbation
theory in the previous Subsection, it is important to get independent
information from further observables that are less sensitive to the
explicit production cross sections from the various channels
employed. We recall that especially the $K^\pm$ yields from
$N\Delta$ and $\Delta \Delta$ channels are model dependent and
cannot directly be measured in experiment. Futhermore, there
are cancellation effects due to the different in-medium potentials for
$K^+$ and $K^-$, which imply that the absolute magnitude of the
spectra alone does not provide stringent information on the
in-medium $K^\pm$ properties.

The collective flow of hadrons is experimentally defined by the
anisotropy in the angular distribution as
\begin{equation}
\label{flow}
\frac{dN}{d \phi}  \sim 1 + 2 v_1 \cos(\phi)+ 2 v_2 \cos(2\phi).
\end{equation}
The coefficients $v_1$ and $v_2$ characterize directed
and elliptic flow, respectively, and can be evaluated from the
transport calculations as
\begin{equation}
\label{vvv}
v_1 = \left. \left\langle {p_x\over p_T }\right\rangle \right|_{y,p_T},
\hspace{2cm} v_2 = \left.\left\langle \frac{p_x^2-p_x^2}{p_x^2+p_y^2}
\right\rangle \right|_{y,p_T}
\end{equation}
where the beam is in $z$-direction and the reaction plane oriented in
$y$-direction. An elliptic flow $v_2 >0$ indicates in-plane emission of
particles, whereas $v_2 <0$ corresponds to a squeeze-out perpendicular
to the reaction plane.

The elliptic flow $v_2$ is very sensitive to the strength of the
interaction of $K^+, K^-$ mesons with the nuclear environment. In the
case of a repulsive potential (as for $K^+$) one expects $v_2<0$, i.e.
a dominant out-of-plane emission of kaons due to the repulsive
interaction with the nucleon spectators. Oppositely, for antikaons -
which are attracted by the nucleon spectators - the coefficient $v_2$
is expected to be positive if the antikaon absorption by nucleon
spectators is not too strong.

Since the deviations from an isotropic angular distribution are small,
one needs a transport calculation with very high statistics to extract
solid numbers for the coefficients $v_1$ (or $\langle p_x \rangle$) and
$v_2$ especially when gating additionally on rapidity $y$ and/or the
transverse momentum $p_T = \sqrt{p_x^2+p_y^2}$.

\begin{figure}[!]
\centerline{\psfig{file=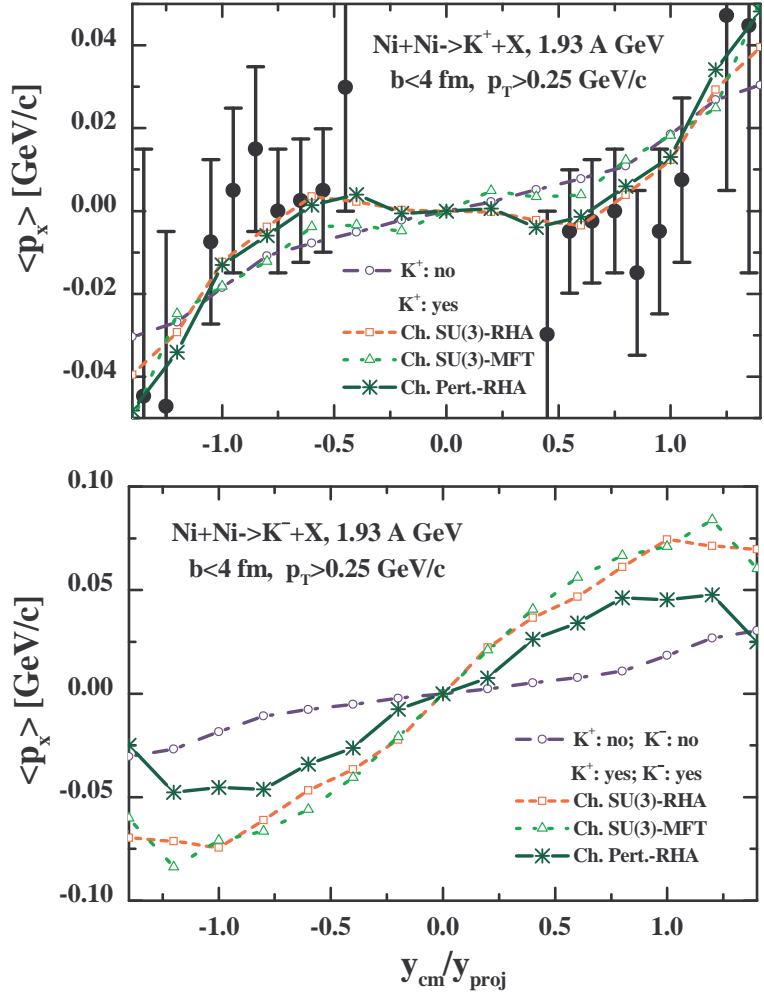,width=10cm}}
\caption{The $\langle p_x \rangle$ for $K^+$ (upper part) and $K^-$ (lower
part) mesons as a function of the normalized rapidity $y_{cm}/y_{proj}$ for
Ni+Ni at 1.93 A$\cdot$GeV. We have gated on central
collisions ($b\le 4$~fm) and applied a transverse momentum cut
$p_T\ge 0.25$ GeV/$c$ as for the experimental data of the FOPI
Collaboration \cite{FOPI_v196} (full dots).
The assignment of the individual lines is the same as in
Fig. \protect\ref{S3_Au15}.}
\label{v1-fopi}
\end{figure}

In Fig. \ref{v1-fopi} we show the $\langle p_x \rangle$ for $K^+$
(upper part) and $K^-$ (lower part) mesons as a function of the
normalized rapidity $y_{cm}/y_{proj}$ for Ni+Ni at 1.93 A$\cdot$GeV. We
have gated on central collisions ($b\le 4$~fm) and applied a transverse
momentum cut $p_T\ge 0.25$ GeV/$c$ as for the experimental data of the
FOPI Collaboration \cite{FOPI_v196} (full dots).
We find that when neglecting any potential for kaons and antikaons the
$K^\pm$ flow follows the proton flow, however, with a lower magnitude.
Since the kaon potential in the chiral SU(3)-MFT model is only very
small, there is almost no change in $\langle p_x \rangle(y)$ (upper
part) for $K^+$ mesons.  Only when including the moderate repulsive
potential of the other models the kaons are 'pushed away' from the
protons and approximately show a vanishing (or even slightly opposite)
flow pattern in line with the data. Consequently, the flow $\langle p_x
\rangle (y)$ leads to the same conclusion on the kaon potential as the
studies in Subsection VI.B.

In contrast, the antikaons are attracted towards the proton flow
direction by an attractive potential. This is most pronounced for the
chiral SU(3)-MFT and -RHA models with their large attraction for
antikaons. The results for the chiral perturbation theory -- including
a moderate $K^-$ attraction -- are in between the 'free' and chiral
SU(3) limits. Unfortunately, there are presently no comparable data to
confirm or exclude the various models.

\begin{figure}[!]
\centerline{\psfig{file=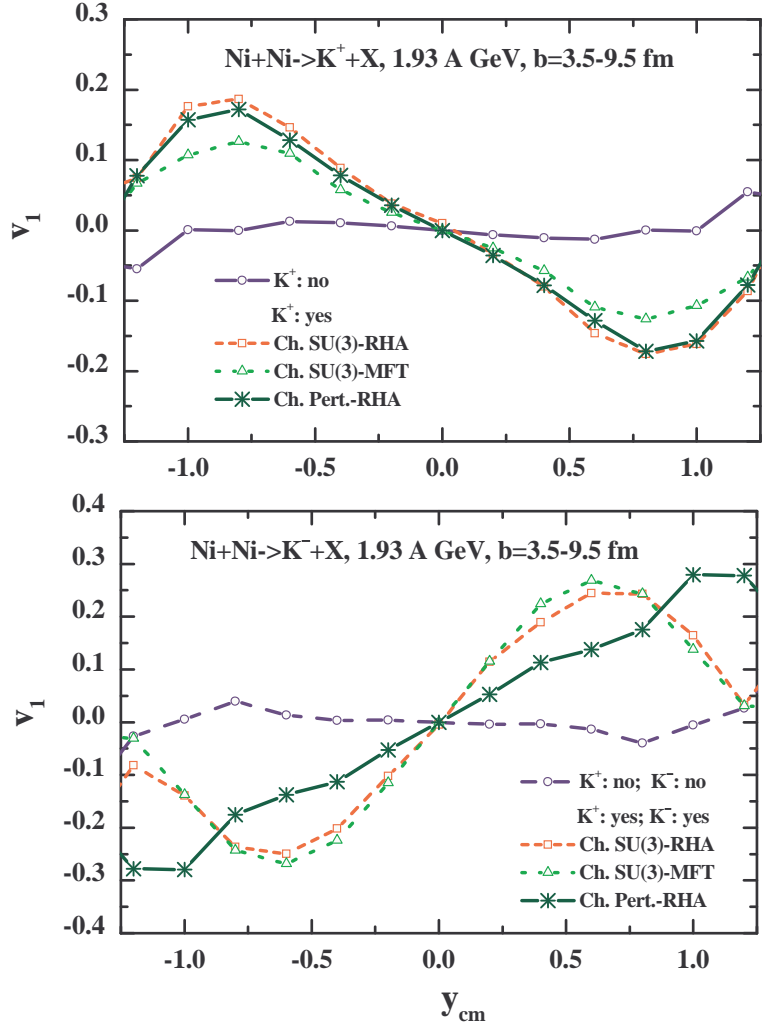,width=10cm}}
\caption{The directed flow $v_1$ for $K^+$ (upper part)
and $K^-$ (lower part) mesons as a function of the
center-of-mass rapidity $y_{cm}$ for
non-central ($b=3.5\div 9.5$~fm) Ni+Ni collisions at 1.93 A$\cdot$GeV.
The assignment of the individual lines is the same as in
Fig. \protect\ref{S3_Au15}.}
\label{v1_y}
\end{figure}

In Fig. \ref{v1_y} we, furthermore, provide predictions for the
directed flow $v_1$ for $K^+$ (upper part) and $K^-$ (lower part)
mesons as a function of the center-of-mass rapidity $y_{cm}$ for
non-central ($b=3.5\div 9.5$~fm) Ni+Ni collisions at 1.93 A$\cdot$GeV.
As it is well known, the flow phenomena are more pronounced for
mid-central and peripheral nucleus-nucleus collisions; this is
also seen from Fig. \ref{v1_y}. Here the strong antiflow of
$K^+$-mesons should allow to further discriminate the models in
the next round of experimental studies. This also holds for the
strong attractive $K^-$ flow in the lower part of Fig. \ref{v1_y}
that needs experimental control.

\begin{figure}[!]
\centerline{\psfig{file=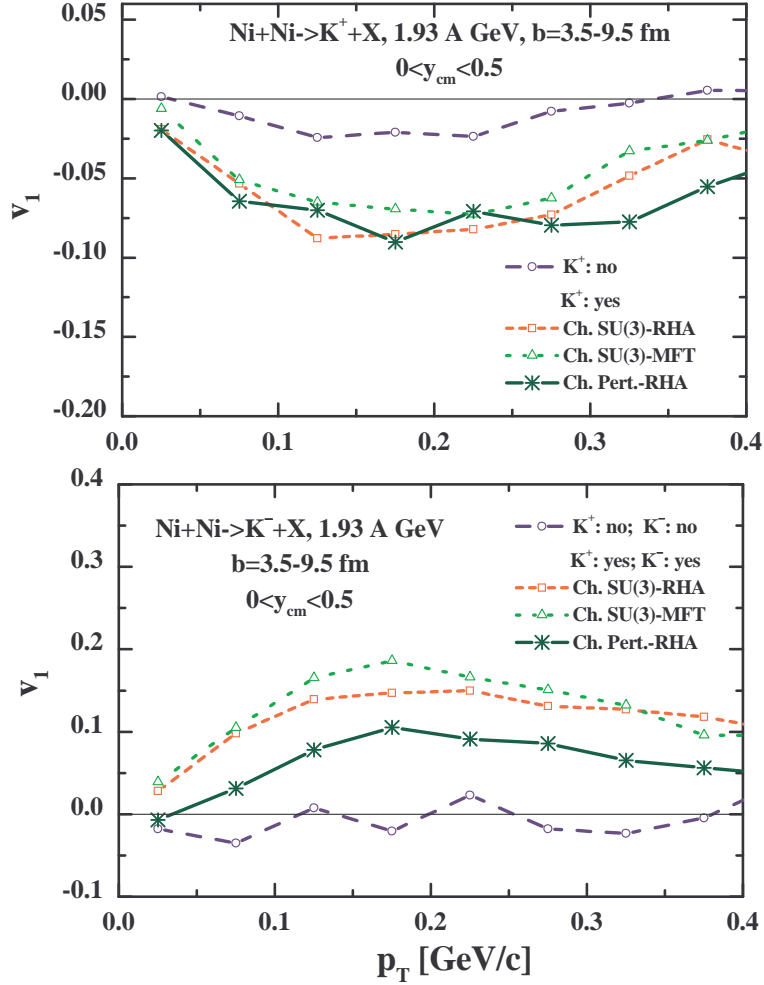,width=10cm}}
\caption{The directed flow $v_1$ for $K^+$ (upper part) and $K^-$
(lower part) mesons as a function of the transverse momentum $p_T$ for
non-central ($b=3.5\div 9.5$~fm) Ni+Ni collisions at 1.93 A$\cdot$GeV
including the rapidity cut $0<y_{cm}<0.5$.  The assignment of the
individual lines is the same as in Fig. \protect\ref{S3_Au15}.}
\label{v1_pt}
\end{figure}

In Fig. \ref{v1_pt} we display the directed flow $v_1$ for $K^+$ (upper
part) and $K^-$ (lower part) mesons as a function of the transverse
momentum $p_T$ for non-central ($b=3.5\div 9.5$~fm) Ni+Ni collisions at
1.93 A$\cdot$GeV including the rapidity cut $0<y_{cm}<0.5$. The
$p_T$-dependence of the $v_1$ is another observable that
will allow for future experimental discrimination. Our calculations
demonstrate that the kaon flow is practically zero if no potentials
are employed. But for nonzero $K^+$ potentials  $v_1$
becomes increasingly negative with $p_T$ up to $p_T\approx 0.4$ GeV/c.
The strength of $v_1$ is, furthermore, almost proportional
to the strength of the kaon self energy. The antikaons show a qualitatively
similar behaviour in $p_T$, but with the opposite sign. There is
practically no signal -- within statistics -- for a non-vanishing
$v_1$ when discarding a $K^-$ potential. Only when
including moderate (for the  chiral perturbation theory) or stronger
antikaon potential (for the chiral SU(3)-RHA and -MFT models) a
positive signal in $v_1$ is found again, which is most
pronounced for momenta in the 0.2 to 0.4 GeV/c regime.

We finally present our results for the elliptic flow $v_2$ for $K^+$
(upper part) and $K^-$ (lower part) mesons as a function of the
center-of-mass rapidity $y_{cm}$ for non-central ($b=3.5\div
9.5$~fm) Ni+Ni collisions at 1.93 A$\cdot$GeV. The calculations
show a distinct dependence of the elliptic flow as a function of
the rapidity in the cms ($y_{cm}$) for kaons as well as antikaons.
For $K^+$ mesons the flow is always negative, i.e. enhanced
perpendicular to the reaction plane. The size of this
'squeeze-out', however, provides a measure for the strength of the
repulsive potential. For free antikaons the elliptic flow is again
compatible with zero (within statistics), but directed in-plane
for all rapidities when including attractive potentials. These
effects are small at midrapidity, but become more pronounced for
$|y_{cm}| > 0.5$, where the spectator nucleons show up.
Also note, that the $K^\pm$ mesons in the interactions with
spectators only probe densities $\rho_B \le \rho_0$.
The next round of experiments should allow to put some further
constraints on the models.

\begin{figure}[!]
\centerline{\psfig{file=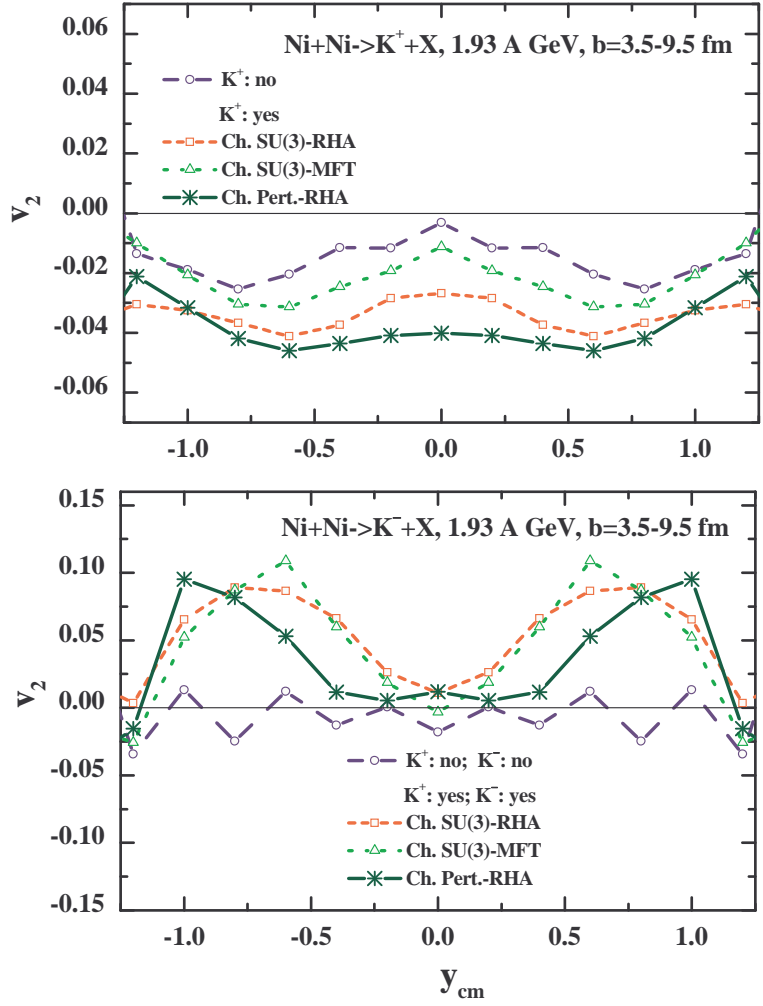,width=10cm}}
\caption{The $v_2$ coefficient for $K^+$ (upper part) and $K^-$ (lower
part) mesons as a function of the center-of-mass rapidity $y_{cm}$ for
non-central ($b=3.5\div 9.5$~fm) Ni+Ni collisions at 1.93 A$\cdot$GeV.
The assignment of the individual lines is the same as in Fig.
\protect\ref{S3_Au15}.}
\label{v2_y}
\end{figure}

We stress, that there is a clear difference for free and in-medium
$K^+, K^-$ scenarios in the $v_1$ (or $\langle p_x \rangle$), $v_2$
flow patterns which provides a unique signal for medium modifications.
Thus, future experiments with sufficient statistics might be able not
only to confirm unambiguously medium effects for $K^+$ and $K^-$
mesons, but also provide constraints on the underlying potentials in a
dense and hot medium.

\section{Summary}

To summarize, we have investigated in a chiral SU(3) model the
temperature and density dependence of the $K, \bar K$-meson masses
arising from the interactions with  nucleons and scalar and vector
mesons. The properties of the light hadrons -- as studied in a SU(3)
chiral model -- modify the $K$-meson properties in the hot and dense
hadronic medium. The SU(3) model with parameters fixed from the
properties of hadron masses, nuclei and hypernuclei and KN scattering
data, takes into account all terms up to the next to leading order
arising in chiral perturbative expansion for the interactions of
$K$-mesons with baryons. The important advantage of the
present approach is that the DN, KN as well as $\pi N$ $\Sigma$ terms
are calculated within the model itself. The predictions for the $\pi N$
and KN $\Sigma$ terms are reasonable, the value for $\Sigma_{KN}$ from
the model being, furthermore, in agreement with lattice gauge
calculations.

Using the Lagrangian from chiral perturbation theory with a
Tomozawa-Weinberg interaction, supplemented by an attractive scalar
interaction (the $\Sigma$ term) for the $KN$
interactions, the results obtained are seen to be similar to earlier
calculations:  the $K^+$ mass increases with density while the $K^-$mass
decreases.  However, the presence of the repulsive range term, given by
the last term in (\ref{ldcpt}), reduces the drop in the antikaon mass.
The chiral effective model, which is adjusted to describe nuclear
properties, gives a larger drop of the $K$-meson masses at finite
density as compared to chiral perturbation theory dominantly due to an
attractive range term. Furthermore, the effect of the baryon Dirac sea
for hot hadronic matter (within the chiral SU(3) model) gives a
slightly higher value for the $K$-meson masses as compared to the
mean-field calculations, while the qualitative trends in the medium remain.

Since the predictions of the various models differ substantially for
the $K^\pm$ masses at finite density and temperature we have used a
covariant transport approach to study the potential effects of the
models in comparison to experimental data at SIS energies.  Our
detailed analysis for $K^\pm$ rapidity and transverse energy
distributions has given a clear preference for the potentials from
chiral perturbation theory, which yields a moderately repulsive kaon
potential and a moderately attractive antikaon potential.
We stress that this moderate $K^-$ potential is in agreement with more
sophisticated coupled-channel calculations, which take into account
effects from the $\Lambda(1405)$ and dress the in-medium $K^-$ propagator
selfconsistently.

We have, furthermore, calculated the collective flow for kaons and
antikaons in Ni+Ni reactions at 1.93 A$\cdot$GeV, which shows a clear
dependence on the strength of the potentials for non-central and
peripheral collisions. Moreover, the dependence of the
flow coefficients $v_1$ and $v_2$ on the transverse momentum $p_T$
and on the rapidity $y$ show a sensitivity to the sign and magnitude of
the $K^\pm$ potentials. These sensitivities can be used to determine
the in-medium properties of the kaons and antikaons from the
experimental side in the near future \cite{Peter04}.

\begin{acknowledgements}

We thank H. Oeschler, J. Reinhardt, P. Senger and L. Tol\'os for
fruitful discussions and W.  Cassing for the actual version of the HSD
transport code used in our analysis.  One of the authors (AM) is
grateful to the Institut f\"ur Theoretische Physik Frankfurt for the
warm hospitality. AM acknowledges financial support from
Bundesministerium f\"ur Bildung und Forschung (BMBF) and ELB from
Deutsche Forschungsgemeinschaft (DFG) and GSI. The support from the
Frankfurt Center for Scientific Computing (CSC) is additionally
gratefully acknowledged.
\end{acknowledgements}


\end{document}